\documentclass[final,5p,times,twocolumn]{elsarticle}

\usepackage{amsmath}
\usepackage{amssymb}
\usepackage{amsfonts}
\usepackage{hyperref}
\usepackage{xcolor}
\usepackage{soul}
\usepackage{booktabs}
\usepackage{multirow}
\usepackage{makecell}
\usepackage{threeparttable}
\biboptions{numbers,sort&compress}
\newcommand{\be}{\begin{equation}}
\newcommand{\ee}{\end{equation}}

\begin{document}

\begin{frontmatter}

\title{Extracting overlapping gravitational-wave signals of Galactic compact binaries: a mini review}

\author[a,b]{Rui Niu}
\affiliation[a]{organization={Department of Astronomy, University of Science and Technology of China, Chinese Academy of Sciences},
            addressline={}, 
            city={Hefei},
            postcode={230026}, 
            state={Anhui},
            country={China}}
\affiliation[b]{organization={School of Astronomy and Space Sciences, University of Science and Technology of China},
            addressline={}, 
            city={Hefei},
            postcode={230026}, 
            state={Anhui},
            country={China}}
            

\author[a,b]{Wen Zhao}
\ead{wzhao7@ustc.edu.cn}

\begin{abstract}
Gravitational wave (GW) observations have provided a novel tool to explore the universe. In the near future, space-borne detectors will further open the window of low-frequency GW band where abundant sources exist and invaluable information for astrophysics, cosmology, and fundamental physics can be revealed.
However, there are various new challenges in data analyses for space-borne detectors coming with the abundance of GW signals.
For example, there are Galactic compact binaries (GCBs) with an overwhelming number that can produce continuous GW signals existing the entire mission time of detectors.
The enormous overlapping GCB signals tangle and correlate with each other, and blend with other types of sources together in the observed data.
Extracting source information from overlapping signals is one of the key problems for data analyses of space-borne detectors.
In the paper, we present a review of currently available solutions for extracting overlapping GCB signals as thoroughly as possible aiming at promoting more interest in this question and inspiring further improvements.
Current solutions can be roughly categorized by two classes, iterative subtraction and global fitting. There are diverse implementations of both strategies with enhancements focusing on different aspects. Meanwhile, the hybrid approach and the machine learning technique are also used in recent years. In the last, we also present an introduction of the stochastic foreground formed by unresolvable faint GCBs about its separation from extra-galactic backgrounds and its utility in exploring the properties of the Galaxy.
\end{abstract}







\end{frontmatter}



\section{Introduction} \label{sec_intro}
Direct detections of gravitational waves (GWs) by ground-based detectors have ushered in a new era of GW astronomy \cite{LVKCollaboration2019,LVKCollaboration2020b,LVKCollaboration2021e,LVKCollaboration2021d}.
Fruitful results have been accomplished based on detected events and researches with GWs are explosively growing in recent years. These detections have initiated a paradigm shift in studies of gravity, astrophysics, and cosmology.
In the near future, space-borne detectors including LISA\cite{AmaroSeoane2017a}, Taiji\cite{Ruan2020}, and TianQin\cite{Luo2016} will open the new window of low-frequency GW band.

One significant difference between current ground-based detectors and future space-borne detectors is that data given by current ground-based detectors are noise-dominant whereas signals are dominant for space-borne detectors. 
For current ground-based detectors, the compact binaries will quickly merge after entering the sensitive band, and the transient signals are sparsely distributed in time \cite{LVKCollaboration2019,LVKCollaboration2020b,LVKCollaboration2021e,LVKCollaboration2021d}. The situations of signal overlapping are expected to be rare. 
{As reported in the work \cite{Relton2021a} where the probability of signal overlapping and how severe overlapping can induce significant bias in parameter estimation for second-generation ground-based detectors are thoroughly investigated, it is unlikely to observe overlapping signals by current existing detectors. }
Whereas, for future space-borne detectors, the enormous sources can persist in the sensitive band during the whole mission period \cite{AmaroSeoane2017a,Ruan2020,Luo2016}. Their GW signals can be heavily overlapping both in the time and frequency domain, which brings new challenges in extracting physical information from the data.
{Similar issues of signal overlapping also arise in third-generation ground-based detectors. Due to the exquisite sensitivities especially the improvement in low frequency, the visible duration and signal number are both significantly increased. It is unlikely to observe signals without overlapping for third-generation detectors, and the events with very close merger time which may suffer from significant bias in parameter estimation can be common \cite{Samajdar2021a,Pizzati2022}. 
There are extensive works about parameter inference techniques and impacts on researches of scientific problems for overlapping signals in third-generation detectors.
For examples, the traditional Bayesian inference framework with strategies of hierarchical subtraction and joint estimation for overlapping signals is investigated in the work \cite{Janquart2022}, and a joint parameter estimation analyzing two overlapping signals simultaneously using normalizing flows is demonstrated in \cite{Langendorff2022}.
In the work \cite{Dang2023}, impacts of signal overlapping on testing General Relativity are elaborated. 
In works \cite{Himemoto2021a,Zhong2023}, the impacts of unresolvable overlapping foreground and subtraction residuals induced by parameter estimation bias due to signal overlapping on detecting cosmological stochastic gravitational wave background with third-generation detectors are investigated.
A review for signal overlapping problems focusing on third-generation ground-based detectors can be found in \cite{Badaracco2024}. In the rest of this paper, we mainly focus on researches targeting space-borne detectors.
}

There are diverse types of sources presented in the space-borne detector sensitive band \cite{AmaroSeoane2023}, such as massive black hole binaries (MBHBs), extreme mass ratio inspirals (EMRIs), and Galactic compact binaries (GCBs), etc.
GCBs are likely to be the most numerous type of GW sources detected in future space GW observation. 
Tens of millions of such sources may exist in the space-borne detector band. Among them, ten of thousands of bright ones are expected to be independently resolvable, while others will constitute a stochastic foreground as confusion noise \cite{Belczynski2010,Ruiter2010,Korol2017,Nelemans2001}. Identification and subtraction of the resolvable GCBs are not only important for extracting other kinds of sources but also can be assistance in researches about stellar and Galactic astrophysics with the advantage that the information carried by GWs will not be affected by crowded matters in the Galaxy \cite{Nelemans2009,Marsh2011,AmaroSeoane2012,AmaroSeoane2012a,Korol2018a,Wilhelm2020}.

Although all sources including MBHBs and EMRIs will be blended together in data-stream from space-borne detectors and people may ultimately need algorithms that can separate or simultaneously fit all different sources, GCBs may be the kind that is most heavily overlapping due to their overwhelming amount and the feature of long-living.
Therefore, the discussions dedicated to GCBs may be the foundation and starting point of the ultimate full algorithms for separating and fitting all overlapping sources. 
In this paper, we will focus on GCBs and present a review of current solutions to tackle enormous overlapping GCB sources as comprehensively as possible, aiming at paving for further improvements in the problems of signal overlapping for data analyses of space-borne detectors.

Current solutions of extracting GCB signals mainly work around the simulated data sets including the earlier Mock LISA Data Challenges (MLDCs) \cite{Arnaud2006,Arnaud2006a,Arnaud2007,Arnaud2007a,Babak2008b,Babak2010,Babak2008a}, the recent resurrected LISA Data Challenges (LDCs) \cite{Baghi2022}, and the Taiji Data Challenges (TDCs) \cite{Ren2023}. These data sets are released to encourage various efforts for tackling unsolved problems in data analyses of space-borne detectors. 
These challenges involve data sets not only dedicated to GCBs, but also blended with various other possible sources, and the complexity has been increasing progressively.
In the MLDC1, isolated GCBs and moderately overlapping signals with dozens of GCBs are concerned \cite{Arnaud2006,Arnaud2006a,Arnaud2007}. The MLDC2 considers the full population with 26 million GCBs, and includes two data sets where the MLDC2.1 only contains signals of GCBs while MLDC2.2 blends signals of GCBs with MBHBs and EMRIs \cite{Arnaud2007a,Babak2008b}.
The data set of GCBs in MLDC3 contains $\sim 60$ million binaries, which is descended from MLDC2.1 with the improvement of realism by considering two different kinds of GCBs, binaries with two detached components and binaries with interacted components \cite{Babak2010,Babak2008a}.
The MLDC4 is the descendant of MLDC2.2, which blends different types of sources concerned separately in sub-challenges of MLDC3 into a single data stream \cite{Babak2010}.

The new LDCs \footnote{\url{https://lisa-ldc.lal.in2p3.fr/}} \cite{Baghi2022} are resumed in recent years with the new design \cite{AmaroSeoane2017a} of LISA. Currently, three challenges have been released, LDC1 \texttt{Radler}, LDC2a \texttt{Sangria}, and LDC2b \texttt{Spritz}.
The LDC1 considers various types of possible LISA sources separately like MLDC3, and includes six sub-challenges where LDC1-4 is dedicated to GCBs and contains 26 million signals.
The LDC2a is the updated challenge similar to MLDC4, which mixes different types of sources.
The LDC2b has the improvement of considering the realistic instrumental and environmental noise, including gaps, glitches, and non-stationary noise. While the GW sources considered in LDC2b are relatively simple, where only MBHBs and verification GCBs are contained.
The Taiji project also releases the data challenge \cite{Ren2023} for the configuration of the Taiji detector, which includes sub-challenges concerning various types of GW sources separately and the mixture of these sources.

Numerous endeavors have been made in previous works to address the problem of signal overlapping.
The ideas for this problem can be roughly categorized into two groups, the iterative subtraction strategy and the global fitting strategy. 
The iterative subtraction strategy searches the maximum likelihood estimation for a single source in each step, and this procedure will be performed iteratively with the remaining data after subtracting the identified signals \cite{Zhang2021a,Zhang2022,Gao2023,Gao2024,Lu2022}.
In contrast, the global fitting strategy fits all signals simultaneously in the full Bayesian approach with sophisticated Markov chain Monte Carlo (MCMC) sampling algorithms to obtain the joint posterior distributions \cite{Umstaetter2005,Umstaetter2005a,Littenberg2020,Littenberg2023,Lackeos2023,Littenberg2024,Karnesis2023,Katz2024}.
Both two strategies are adopted and implemented in diverse full-scale and end-to-end pipelines for extracting overlapping signals of GCBs.
{The iterative subtraction solutions can extract source candidates quickly but suffer from the correlations among overlapping signals and the contamination of accumulating signal residuals left by each subtraction iteration. 
The global fitting solutions employ the full Bayesian approach to analyze all sources in a band simultaneously, which can better deal with the source correlations and residual contamination, but with the price of extremely massive demand for computational resources.}
The hybrid approach combining the maximum likelihood estimation and the Bayesian parameter estimation with MCMC is also proposed for combining the strengths of the two strategies while evading their drawbacks \cite{Strub2024,Strub2022,Strub2023}.
{The flourishing machine learning algorithms provide new avenues to solve parameter estimation problems. 
After training with simulated data, the algorithms can generate posteriors directly without enormously evaluating the computationally expensive likelihood, which can complete the parameter estimation nearly in real time. 
The techniques have been successfully used in the parameter estimation for current ground-based detector data \cite{Dax2021,Wildberger2023,Dax2023,Dax2021b,Green2020a,Green2020}.
}
Machine learning techniques are also considered in the problem of overlapping GCBs, which are expected to be prospective powerful tools in future data analyses of space-borne detectors \cite{Korsakova2024}.
The resolvable sources are only a small fraction of the whole GCB population, the remaining faint GCBs can form a stochastic foreground. On the one hand, this Galactic foreground plays the role of noise deteriorating the signal-to-noise ratio (SNR) of individual sources and blending with stochastic signals from extra-galactic sources \cite{Seto2004,Robson2017,Wu2023a,Liu2023,Flauger2021,Adams2010,Adams2014,Banagiri2021,Boileau2021,Poletti2021}. On the other hand, this foreground also contains information on the GCB population which is correlated to properties of the Galaxy, and offers a unique tool to study Galactic astrophysics \cite{Breivik2020a,Benacquista2006,Georgousi2022}.

The rest of this paper is organized as follows. In the next section, we briefly introduce the characterizations of GW signals from GCBs and how detectors respond to them. 
Section \ref{sec_solutions} is the main part of this paper where we present a detailed review of efforts on extracting overlapping GCB signals reported in recent years.
After briefly summarizing earlier works around MLDC in Section \ref{subsec_early}, we first introduce a typical implementation of the iterative subtraction scheme and two variants focusing respectively on reducing inaccurate subtraction contamination and improving search efficiency in Section \ref{subsec_iterative}. 
Then, we elucidate basic conceptions of the global fitting scheme and introduce two independent implementations of this strategy in Section \ref{subsec_global}.
The hybrid Bayesian approach combining the maximum likelihood estimation and the MCMC sampling, and a preliminary attempt of utilizing machine learning techniques in solving the problem of overlapping GCB signals are introduced next in Section \ref{subsec_hybrid} and \ref{subsec_ml}.
The discussions about unresolvable GCBs are given in Section \ref{sec_unresolvable}.
The final summary is presented in Section. \ref{sec_summary}.

\section{Galactic compact binaries} \label{sec_GCBs}
\begin{figure}
    \centering
    \includegraphics[width=\columnwidth]{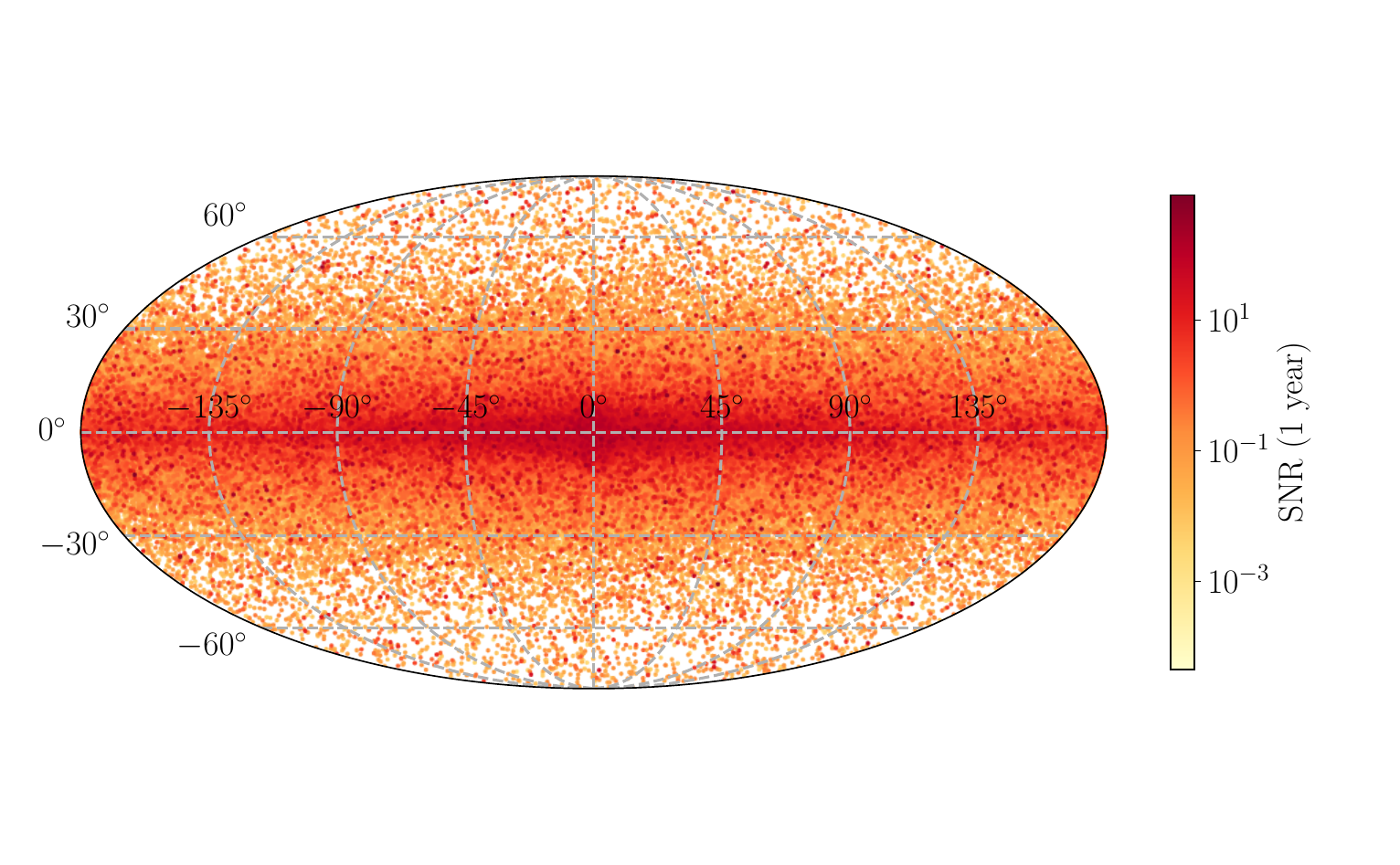}
    \caption{\textbf{Illustration for sky distribution of GCBs}. 
    The simulated GCB catalog is from LDC2a \cite{Baghi2022}. The SNRs are calculated with 1-year observation and the sources are displayed in the Galactic coordinate.}
    \label{fig_sky}
\end{figure}

According to population models \cite{Belczynski2010,Ruiter2010,Korol2017,Nelemans2001}, there are tens of millions of compact binaries in the Galaxy that are slowly inspiraling towards each other with emissions of GWs in the mHz band and might be the type of most numerous sources observed by the space-borne detectors. 
An illustration of the sky distribution of a simulated GCB population taken from the training dataset of LDC2a \cite{Baghi2022} is presented in Figure \ref{fig_sky}.
Among the population, tens of thousands of GCBs are expected to be individually resolvable through the four-year observation time of space-borne detectors.
Current electromagnetic observations have identified about dozens of GCBs\footnote{\url{https://gitlab.in2p3.fr/LISA/lisa-verification-binaries}} \cite{Kupfer2024}, while hundreds are predicted to be detected by future observations \cite{Korol2017}. 
These known sources are referred to as verification binaries and the loud ones are guaranteed to be detectable by space-borne GW detectors.
The loud verification binaries are expected to be quickly identified by just weeks integration time of observation, thus offering an important tool for functional tests and performance monitoring of the instruments \cite{Kupfer2018}. 
Most GCBs are binary white dwarfs including detached binaries, as well as {interacted binaries that have reached the Roche lobe overflow and started mass transfer \cite{Nissanke2012}.} While a small fraction of GCBs may involve with neutron stars or black holes. A summary of expectation of GCBs in the space-borne detector band is present in Table \ref{tab_nums} which is cited from \cite{AmaroSeoane2023}. 
More details about population synthesis simulations or formation scenarios can be found in reviews \cite{Benacquista2022,AmaroSeoane2023}.


GCBs in the band of space-borne detectors are far from merger, and will stay in the inspiral phase of slow chirping during the entire observation period.
GWs radiated from GCBs can be well described by the quasi-monochromatic waveform which takes the form of \cite{Katz2022b,Cornish2007}
\begin{equation} \label{eq_hchp}
    \begin{aligned}
        h_+(t) & = \mathcal{A} (1+\cos^2 \iota) \cos \Phi(t), \\
        h_\times(t) & = -2 \mathcal{A} \cos \iota \sin \Phi(t).
    \end{aligned}
\end{equation}
Here $\mathcal{A}$ is the amplitude, $\iota$ is the inclination of binary orbit, and $\Phi$ is the GW phase which can be expressed as $\Phi(t) = \phi_0 + 2\pi\int f(t') dt'$ with the arbitrary initial phase $\phi_0$ and the frequency evolution $f(t)$.
Since the frequency evolution is extremely slow for GCBs in the early inspiral stage, $f(t)$ can be characterized by the central frequency $f_0$ and the first derivative $\dot{f}$.
The evolution of amplitude is usually neglected and $\mathcal{A}$ is considered as a constant.
The effects of mass transfer between binaries can be encoded into the parameter $\dot{f}$ \cite{Wang2023c}.
The phase evolution can be written as 
\begin{equation} \label{eq_phase}
    \Phi(t) = 2\pi f_0 t + \pi \dot{f} t^2 + \phi_0.
\end{equation}
The total matric perturbation can be assembled by the sum of two polarizations as
\begin{equation}
    \boldsymbol{h}^{\text{TT}} = \boldsymbol{\epsilon}_+ h_+ + \boldsymbol{\epsilon}_\times h_\times,
\end{equation}
where $\boldsymbol{\epsilon}_+$ and $\boldsymbol{\epsilon}_\times$ denote the polarization tensors which depend on the source location $(\beta, \lambda)$ and the polarization angle $\psi$.
The response to GWs of a single laser link of space-borne detectors is given by \cite{Cornish2003,Rubbo2004,Krolak2004}
\begin{equation}
    y_{sr}= \frac{1}{2(1-\hat{\boldsymbol{k}} \cdot \hat{\boldsymbol{n}}_{sr})}
    \hat{\boldsymbol{n}}_{sr} 
    \cdot [\boldsymbol{h}^{\text{TT}}(t-L-\hat{\boldsymbol{k}}\cdot\boldsymbol{p}_s) - \boldsymbol{h}^{\text{TT}}(t-\hat{\boldsymbol{k}}\cdot\boldsymbol{p}_r) ]\cdot
    \hat{\boldsymbol{n}}_{sr},
\end{equation}
where the subscripts $s, r$ denote the laser sender node and the receiver node respectively, $\hat{\boldsymbol{k}}$ denotes the unit vector of the GW propagation direction, $\hat{\boldsymbol{n}}_{sr}$ denotes the unit vector along the direction of the link, $\boldsymbol{p}_s$ and $\boldsymbol{p}_r$ are the vectors of positions of the sender node and receiver node in the heliocentric coordinate.
Together with the parameters describing GW strains of GCBs as shown in Equation \ref{eq_hchp} and \ref{eq_phase}, there are 8 parameters $(\mathcal{A}, f_0, \dot{f}, \iota, \lambda, \beta, \psi, \phi_0)$ to fully characterize a signal from GCB.

Since the noise behavior can be conveniently characterized by the power spectral density (PSD) in the frequency domain if the noise is stationary and Gaussian, data analyses of GWs are often performed in the frequency domain.
The method proposed in \cite{Cornish2007} can compute the Fourier transformation of the response quickly and accurately by heterodyning the response signal with a carrier wave of the frequency $f_0$.
By multiplying with the carrier wave, the response signal can be decomposed into the slow part and the fast part.
The Fourier transformation of the fast part can be obtained analytically.
The slow part is transformed through fast Fourier transformation numerically, whereas the number of time samples is significantly reduced.
For a signal extending within the band of $[f_0, (1+\eta)f_0]$, the heterodyning operation can shift the required Nyquist frequency from $2(1+\eta)f_0$ to $2\eta f_0$. Since the GCBs signals are quasi-monochromatic with $\eta \sim 10^{-6}$ \cite{Nelemans2001,Marsat2018}, the number of samples can be much less than the original time samples when numerically computing the fast Fourier transformation for the slow part.
There is open source code \texttt{GBGPU} \cite{Katz2022c,Katz2022b} that can be used to obtain the response signals of GBCs in practice.

\begin{figure}
    \centering
    \includegraphics[width=\columnwidth]{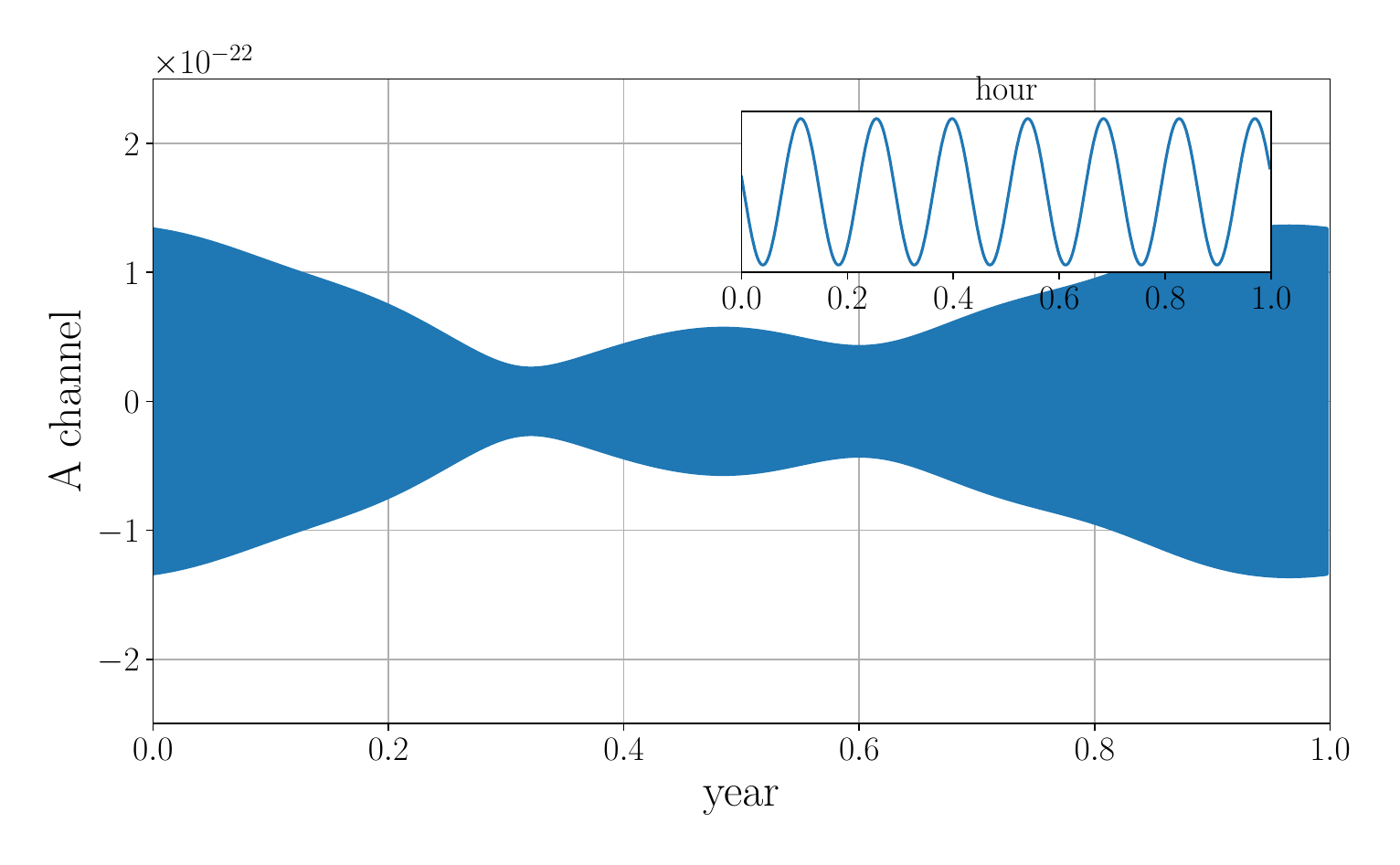}
    \caption{\textbf{Time domain signal of a typical GCB source.} In the mHz band, the GCBs are at the early inspiral stage of their orbital evolution where the radiated GWs are quasi-monochromatic. However, the motion of detectors can endow signals with annual modulation which depends on the sky locations of the sources. 
    The signal shown here is the A channel of 1.5th generation TDI combination for the responses of LISA to a verification source {AM CVn} \cite{Kupfer2024}. The signal is generated by \texttt{lisaanalysistools} \cite{Katz2024a} and \texttt{fastlisaresponse} \cite{Katz2022a}.}
    \label{fig_td}
\end{figure}

\begin{figure}
    \centering
    \includegraphics[width=\columnwidth]{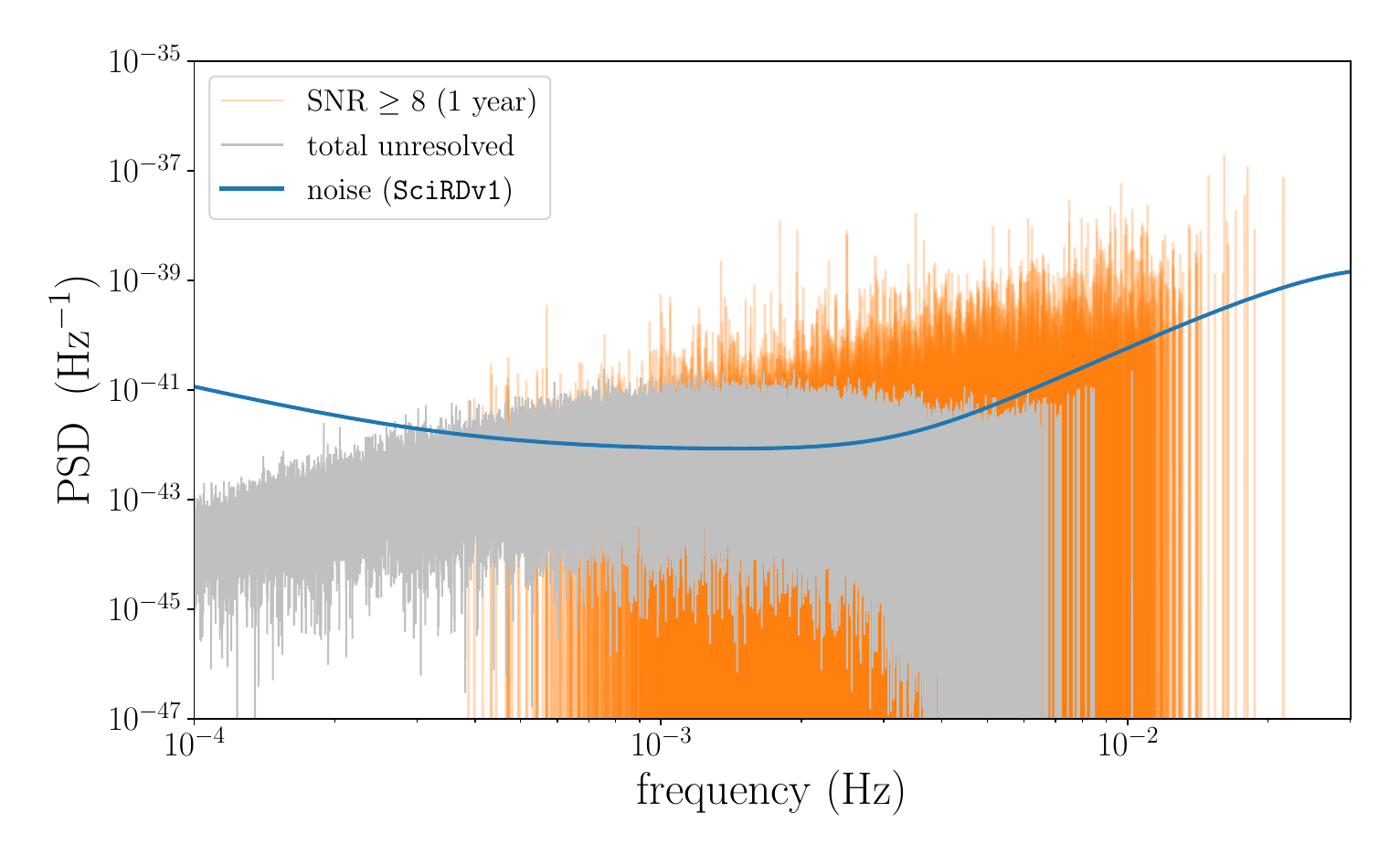}
    \caption{\textbf{Signals of the simulated GCB population in frequency domain.} The GCB catalog is from the training dataset of LDC2a \cite{Baghi2022}. The displayed signals are responses in the A channel of 1.5th generation TDI combination of LISA generated by \texttt{gbgpu} \cite{Katz2022b,Katz2022d}. The orange lines denote the sources that have SNR $\geq$ 8 for 1-year observation and are expected to be resolvable individually. The total contribution of remaining faint sources is denoted by gray lines which can form a stochastic foreground as confusion noise. The noise of the instrument is indicated by the blue line, which is given by the noise model \texttt{SciRDv1} \cite{LISA2018,Babak2021}.}
    \label{fig_fd}
\end{figure}

The space-borne detectors are unequal arm interferometers where the laser frequency noise will experience different time delays when traveling along different arms and cannot be canceled out by itself at the photodetector like ground-based detectors.
In order to suppress the laser frequency noise which can be stronger than GW signals a few orders, the technique called time delay interferometer (TDI) where the observables are created by time-shifting and combining single link responses has to be used to construct artificial equal arm interferometers \cite{Tinto2020,Krolak2004,Vallisneri2005}.
The 1.5th generation (or 1st generation in some literatures) TDI observable $X$ is given by \citep{Babak2021}
\begin{equation}
\begin{aligned}
    X_{1.5} ={} & y_{13} + D_{13}y_{31} + D_{13}D_{31}y_{12} + D_{13}D_{31}D_{12}y_{21} - \\
    &y_{12} - D_{12}y_{21} - D_{12}D_{21}y_{13} - D_{12}D_{21}D_{13}y_{31},
\end{aligned}
\end{equation}
{where $D_{ij}$ denotes the delay operator defined as $D_{ij}y_{sr} = y_{sr}(t-L_{ij})$ and $L_{ij}$ is the arm-length between the node $i$ and node $j$ of the constellation. (In the construction of TDI combinations, the non-commutativity and variations of arm-length may be taken into consideration, whereas when actually computing the detector responses, the approximation of rigid constellation where all arm-length is equal and constant is usually adopted \cite{Babak2021,Zhao2024b,Marsat2018}. Therefore, we follow this convention and use $L_{ij}$ to denote the arm-length when writing TDI combinations, while using $L$ in other places.)}
The other two observables $Y, Z$ can be obtained similarly by cyclic permutation of indices.
The 2nd generation TDI incorporates that the delay operators are non-commutative for forward and inverse delay of a link due to the rotation of the constellation by compensating more virtual loops in two arms as \citep{Babak2021}
\begin{equation}
\begin{aligned}
    X_{2.0} ={} 
    & y_{13} + D_{13}y_{31} + D_{13}D_{31}y_{12} + D_{13}D_{31}D_{12}y_{21} + \\
    & D_{13}D_{31}D_{12}D_{21}y_{12} + D_{13}D_{31}D_{12}D_{21}D_{12}y_{21} + \\
    & D_{13}D_{31}D_{12}D_{21}D_{12}D_{21}y_{13} + D_{13}D_{31}D_{12}D_{21}D_{12}D_{21}D_{13}y_{31} - \\
    & y_{12} - D_{12}y_{21} - D_{12}D_{21}y_{13} - D_{12}D_{21}D_{13}y_{31} - \\
    & D_{12}D_{21}D_{13}D_{31}y_{13} - D_{12}D_{21}D_{13}D_{31}D_{13}y_{31} - \\
    & D_{12}D_{21}D_{13}D_{31}D_{13}D_{31}y_{12} - D_{12}D_{21}D_{13}D_{31}D_{13}D_{31}D_{12}y_{21}. 
\end{aligned}
\end{equation}
The constructions of TDI observable are not unique and have abundant forms. Different constructions can have different sensitivities to GWs of different polarizations and propagation directions \cite{Wang2023d,Wang2022c}.
The $X, Y, Z$ channels are correlated, and the independent TDI channels can be obtained through the combination of 
\begin{equation}
    \begin{aligned}
        A &= \frac{1}{\sqrt{2}}(Z - X), \\
        E &= \frac{1}{\sqrt{6}}(X - 2Y + Z), \\
        T &= \frac{1}{\sqrt{3}}(X + Y + Z). \\
    \end{aligned}
\end{equation}
In the low-frequency limit where $f<(1/2\pi L)$ {with $L$ denoting the arm-length of the detector,} signals in the $A$ channel are mainly contributed by the plus polarization, the $E$ channel can be approximated to the cross polarization, and the $T$ channel is approximated to the breath polarization but which is absent in General Relativity. Thus, in data analyses with low-frequency approximation, only $A$ and $E$ channels are considered usually \cite{Cornish2020a,Marsat2021,Strub2022}.
We illustrate the time domain signal in the $A$ channel of a typical GCB in Figure \ref{fig_td} and the frequency domain signals of the GCB population in Figure \ref{fig_fd}.

\begin{table}
    \centering
    \begin{tabular}{lll}
        \toprule
        Types & $N_{\text{total}}$ & $N_{\text{detected}}$  \\
        \midrule
        WD+WD & $\sim 10^8$        & $6000-10000$          \\
        NS+WD & $\sim 10^7$        & $100-300$             \\ 
        BH+WD & $\sim 10^6$        & $0-3$                 \\
        NS+NS & $\sim 10^5$        & $2-100$               \\
        BH+NS & $\sim 10^4-10^5$   & $0-20$                \\
        BH+BH & $\sim 10^6$        & $0-70$                \\\bottomrule
    \end{tabular}
    \caption{\textbf{Expected numbers for the total and detectable sources of various GCB types in the Galaxy.} This table is cited from \cite{AmaroSeoane2023}.
    }
    \label{tab_nums}
\end{table}

\section{Solutions for extracting overlapping GCB signals} \label{sec_solutions}
This section provides a detailed review of currently available solutions for extracting overlapping GCB signals.
We mainly focus on recent works, early efforts with MLDCs are briefly summarized in Section \ref{subsec_early}, and diverse innovations and new implementations in recent years are introduced in the subsequent sections.
A summary of solutions mentioned here and corresponding references are presented in Table \ref{tab_ref_summary}.

\subsection{Early researches} \label{subsec_early}
Problems of extracting overlapping GCB signals have been discussed for more than two decades.
The iterative subtraction strategy was proposed at the beginning of this century \cite{Cornish2003a}, where the brightest sources are iteratively identified and subtracted.
The global fitting strategy can be traced back to the early researches \cite{Umstaetter2005a,Umstaetter2005} which employs the trans-dimensional MCMC sampling algorithm to simultaneously infer the source number, joint posteriors of source parameters, and noise levels.

Diverse solutions for overlapping GCB signals are explosively presented working around the MLDCs.
For example, the blocked annealed metropolis (BAM) algorithm developed in \cite{Cornish2005,Crowder2006,Crowder2007} is a quasi-Bayesian approach with a $\mathcal{F}$-statistic likelihood and a customized MCMC sampling strategy. The subsequent work \cite{Littenberg2011} extends the BAM algorithm through introducing parallel tempering, the reversible jump MCMC (RJMCMC), and the fast-slow decomposition waveform model, which is the basis of the recent full-scale global fitting pipeline GBMCMC introduced in Section \ref{subsec_global}.
The search methods using $\mathcal{F}$-statistic and various optimal algorithms \cite{Rubbo2006,Whelan2010,Whelan2008,Prix2007,Blaut2010,Bouffanais2016} are also the foundation of diverse iterative subtraction solutions in recent years.
Additionally, various intelligent ideas besides the above two schemes are also widely discussed, including a tomographic approach \cite{Mohanty2006,Nayak2007}, genetic searches \cite{Crowder2006a}, the two-stage strategy \cite{Stroeer2007}, and a sophisticated MCMC walking strategy \cite{Trias2009,Trias2008b}, etc.
More details about early efforts can be found in reports of each MLDC \cite{Arnaud2007,Babak2008b,Babak2010,Babak2008a}, and a comprehensive review \cite{Vallisneri2009}.

Based on early efforts, various enhancements, new methods and implementations have emerged with the resurrected LDCs in recent years, which will be detailed in the following.

\subsection{Iterative substraction using $\mathcal{F}$-statistic and particle swarm optimization} \label{subsec_iterative}
One of widely considered strategies for extracting overlapping signals is iterative subtraction where the search algorithm for one individual source will be run to identify the brightest source, then the identified signal will be subtracted from the data, and this procedure will be iteratively executed with the residual data until some stopping criteria are satisfied. 
The idea of iterative identification and subtraction for overlapping signals in mHz GW band is proposed early in \cite{Cornish2003a}, and is implemented extensively in previous works.
In this review, we present an introduction of a typical iterative subtraction scheme developed in recent years \cite{Zhang2022,Zhang2021a,Gao2023}, which employs $\mathcal{F}$-statistic to construct likelihood and uses particle swarm optimization (PSO) to search the optimal estimations.

The framework developed in \cite{Zhang2021a} is referred to as Galactic Binary Separation by Iterative Extraction and Validation using Extended Range (GBSIEVER). In the iterative subtraction scheme, the single brightest source needs to be identified in each iteration. GBSIEVER implements this through maximum likelihood estimation where the likelihood is constructed by $\mathcal{F}$-statistic.
The data from detectors are composed of noise $n(t)$ and GW signals $h(t)$, which can be expressed as $d(t) = n(t) + h(t)$.
Assuming the noise $n(t)$ is Gaussian and stationary, the probability of a realization of $d(t)$ given a specific GW signal $h(t)$ described by a set of parameters $\boldsymbol{\theta}$ can be written by
\begin{equation} \label{eq_likelihood}
    \ln p(d(t) | h(t, \boldsymbol{\theta})) = -\frac{1}{2} \langle d(t)-h(t, \boldsymbol{\theta}),\, d(t)-h(t, \boldsymbol{\theta})\rangle, 
\end{equation}
where the angle brackets denote the noise weighted inner product defined as 
\begin{equation} \label{eq_inner_product}
    \langle a(t), \, b(t)\rangle = 4 \mathrm{Re} \int_{0}^{\infty} \frac{\tilde{a}(f) \tilde{b}^*(f)}{S_n(f)}\,df.
\end{equation}
Here, $\tilde{a}(f)$ and $\tilde{b}(f)$ is the Fourier transformation of time series $a(t)$ and $b(t)$, $S_n(f)$ denotes the PSD of the noise.
To identify the signal in data through maximum likelihood estimation, one needs to explore the parameter space of $\boldsymbol{\theta}$ to find the parameters $\hat{\boldsymbol{\theta}}_{\mathrm{MLE}}$ where the probability (Equation \ref{eq_likelihood}) has the maximum value.
Drop the term independent with the source parameters, the likelihood can be written by 
\begin{equation} \label{eq_lnLambda}
    \ln \Lambda = \langle d, \, h\rangle - \frac{1}{2}\langle h, \, h\rangle.
\end{equation}
For multiple independent measurements of the same signal, such as the independent TDI channels or different detectors, the final log-likelihood is the sum of log-likelihoods for each independent measurement. The estimation of source parameters $\hat{\boldsymbol{\theta}}_{\mathrm{MLE}}$
is given by
\begin{equation} \label{eq_maxL}
    \frac{\partial\ln \Lambda}{\partial \boldsymbol{\theta}}\bigg|_{\boldsymbol{\theta}=\hat{\boldsymbol{\theta}}_{\mathrm{MLE}}}  = 0.
\end{equation}
Here, we use the hat to denote parameters with the maximum likelihood value.

Exploring the high-dimension parameter space to solve Equation \ref{eq_maxL} is computationally expensive. In the framework GBSIEVER, the reduced likelihood is constructed through $\mathcal{F}$-statistic, where only four intrinsic parameters are searched and the extrinsic parameters can be obtained analytically.
The full set of parameters to describe GCB signals is separated into intrinsic parameters $\{f_0, \dot{f}, \lambda, \beta\}$ and extrinsic parameters $\{\mathcal{A}, \iota, \psi, \phi_0\}$. 
The GW signal can be decomposed by a linear combination of templates only depending on intrinsic parameters as
\begin{equation}
    h(t, \boldsymbol{\theta}) = \Sigma_i \, a_i X_i(t, \boldsymbol{\kappa}),
\end{equation}
where $\boldsymbol{\kappa}$ denotes a set of intrinsic parameters, $a_i$ is a reparametrization of extrinsic parameters, and basis templates $X_i(t, \boldsymbol{\kappa})$ can be defined by the GW signals at 4 sets of specific extrinsic parameters. 
Using this decomposition, we can construct the vector $\boldsymbol{U}$ which has the element of
\begin{equation}
    U_i = \langle d(t) ,\, X_i(t, \boldsymbol{\kappa}) \rangle,
\end{equation}
and the matrix $\boldsymbol{W}$ which has the element of
\begin{equation}
    W_{ij} = \langle X_i(t, \boldsymbol{\kappa}) ,\, X_j(t, \boldsymbol{\kappa}) \rangle.
\end{equation}
Using vector $\boldsymbol{U}$ and matrix $\boldsymbol{W}$ together with the vector $\boldsymbol{A}$ consisting of $a_i$, the likelihood Equation \ref{eq_lnLambda} can be rewritten as 
\begin{equation}
    \ln \Lambda = \boldsymbol{A}\boldsymbol{U} - \frac{1}{2} \boldsymbol{A}\boldsymbol{W}\boldsymbol{A}^T.
\end{equation}
Maximizing the original likelihood $\ln \Lambda$ can be substituted by maximizing the $\mathcal{F}$-statistic defined as 
\begin{equation}
    \mathcal{F}(\boldsymbol{\kappa}) = \boldsymbol{U}^T \boldsymbol{W}^{-1} \boldsymbol{U},
\end{equation}
from which one can obtain the maximum likelihood estimation for intrinsic parameters $\hat{\boldsymbol{\kappa}}$.
The estimation of $\hat{\boldsymbol{a}}$ can be obtained analytically through
\begin{equation}
    \boldsymbol{A}^T = \boldsymbol{W}^{-1} \boldsymbol{U}.
\end{equation}
The full set of parameters with the maximum likelihood can be recovered through $\hat{\boldsymbol{\kappa}}$ and $\hat{\boldsymbol{a}}$.

The remaining question is how to efficiently explore the intrinsic parameter space to maximize $\mathcal{F}(\boldsymbol{\kappa})$. In the framework of GBSIEVER, PSO is performed to accomplish this task.
PSO is proposed by Kennedy and Eberhart \cite{Kennedy1995} and widely used in various optimal problems.
PSO is inspired by the social behavior of organisms like bird flocks or fish schools.
It utilizes a population (referred to as the swarm) of candidate solutions (referred to as particles) which are moved by the guidance of own experience of each particle and the collective knowledge of the entire swarm to find the optimal solution in the parameter space.
The interaction among particles endows the PSO method with more capability for global searching and against trapping in local optima.

Initially, the particles are randomly drawn in the parameter space and assigned with random velocities. The subsequent movement of particles is guided by \cite{Gao2024}
\begin{equation} \label{eq_pso_move}
    \begin{aligned}
        \boldsymbol{v}_i^{t+1} &= \omega  \boldsymbol{v}_i^{t} + c_1 \boldsymbol{r}_1 \left( \boldsymbol{P}_i^t - \boldsymbol{x}_i^t\right) + c_2 \boldsymbol{r}_2 \left( \boldsymbol{G}^t - \boldsymbol{x}_i^t\right), \\
        \boldsymbol{x}_i^{t+1} &= \boldsymbol{x}_i^{t} + \boldsymbol{v}_i^{t+1}.
    \end{aligned}
\end{equation}
In the above equations, $\boldsymbol{x}_i^t$ and $\boldsymbol{v}_i^t$ denote the position and velocity of $i$-th particle at the iteration step of $t$. $\boldsymbol{P}_i^t$ is the best position of the $i$-th particle found in previous iterations, which represents the experience of individual particles, and $\boldsymbol{G}^t$ is the best position found by the entire swarm in previous iterations, which represents the collective knowledge of the swarm. The terms corresponding to $\boldsymbol{P}_i^t$ and $\boldsymbol{G}^t$ are usually called the personal term and the social term respectively in literature. 
$\omega$, $c_1$, and $c_2$ are constant coefficients that need to be tuned according to the specific problem through experimental runs or empirical knowledge.

The process of optimization can be viewed as two phases, the exploration phase where the particles explore the parameter space expansively and quickly to find better locations, and the exploitation phase where the particles have converged within a promising region and updates of better position will be relatively slow.
$\omega$ is called the inertia weight controlling the balance of exploration and exploitation in the optimization process.
A lower inertia weight favors exploitation and allows particles to quickly converge toward promising regions.
Conversely, a higher inertia weight can help particles resist attractions of previous best positions of personal and social terms, explore the parameter space more expansively, and potentially escape from local optima.
The coefficients $c_1$ and $c_2$ are called acceleration coefficients controlling the balance between the personal term and social term of the particles in their movement.
The setting of these coefficients has significant impacts on the performance and efficiency of PSO.
A typical configuration in the context of GW data analyses can be found in \cite{Normandin2018}, which is also adopted by GBSIEVER.
After sufficient iterations, the particles are expected to converge to the optimal solution.

Some practical techniques are used in GBSIEVER to improve search efficiency and eliminate false candidates.
As introduced in Section \ref{sec_GCBs}, GCB signals have the characterization of narrowband.
Therefore, in practice, it is usually to divide the whole frequency band into small bins.
Independent analyses are performed in different frequency bins, and the edge effects are carefully addressed in the meanwhile.
Furthermore, another practical technique in GBSIEVER is a special downsampling operation which can reduce the number of samples thus relieve the computational cost in likelihood evaluation.
To eliminate spurious sources, GBSIEVER employs a cross-validation procedure to compare the extracted candidates in two independent runs. Only the candidates with similar estimation results in two independent runs are considered to be genuine.

\begin{figure}
    \centering
    \includegraphics[width=\columnwidth]{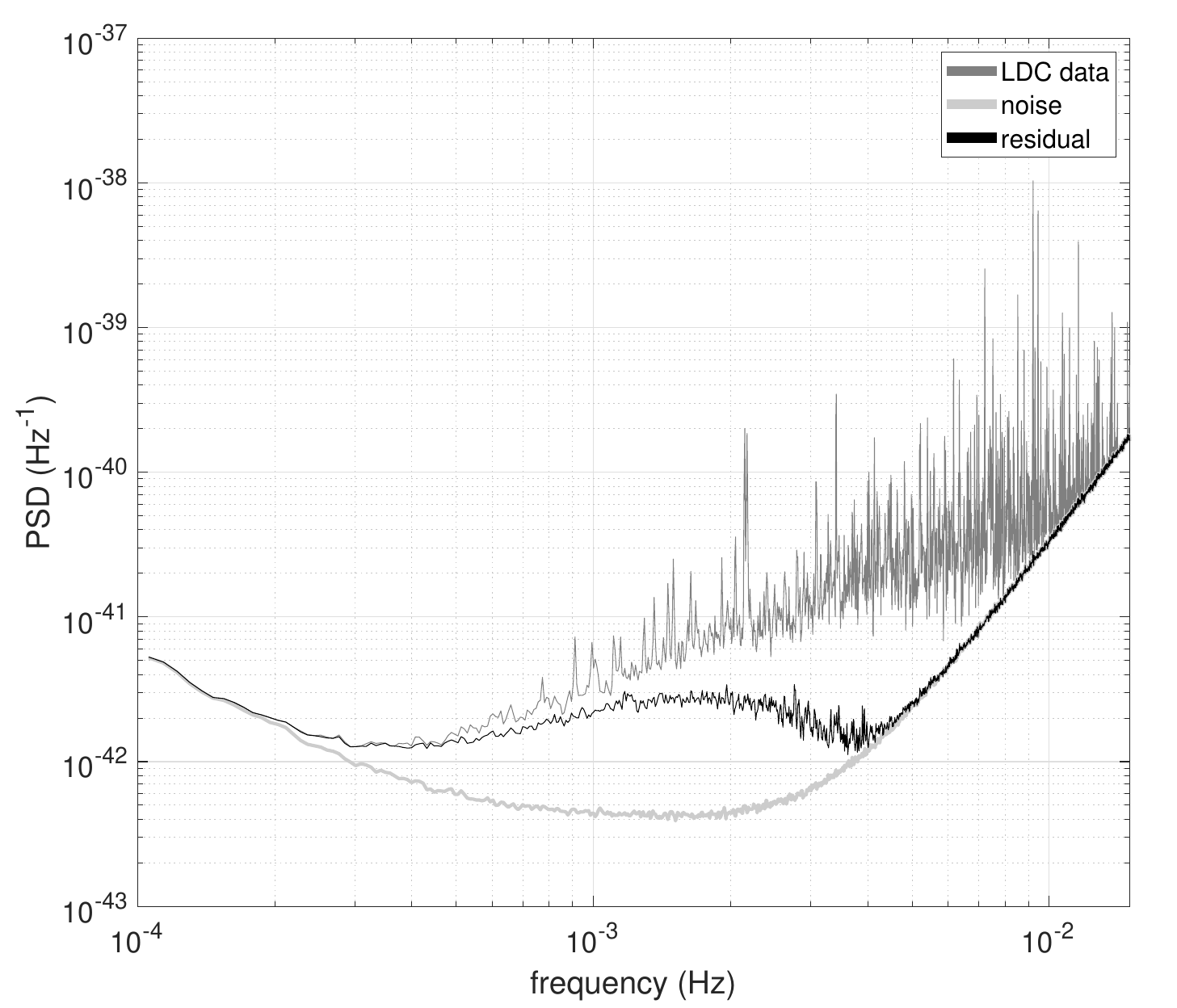}
    \caption{\textbf{The residual after subtracting identified signals given by GBSIEVER. }
    Utilizing the iterative subtraction method with the $\mathcal{F}$-statistic likelihood and the PSO algorithm, GBSIEVER can successfully identify $\mathcal{O}(10^4)$ GCB signals. This figure shows the residual of the A channel of the TDI combination of simulated LISA dataset LDC1-4 which is used to verify the performance of GBSIEVER. This figure is cited from \cite{Zhang2021a}.}
    \label{fig_GBSIEVER_residual}
\end{figure}

The performance of GBSIEVER is verified with the dataset of LDC1-4 and modified MLDC3.1, where it is demonstrated that $\mathcal{O}(10^4)$ GCBs can be successfully identified \cite{Zhang2021a}. 
The residual after subtracting identified signals is shown in Figure \ref{fig_GBSIEVER_residual}.
In the subsequent work \cite{Zhang2022}, the framework GBSIEVER is extended by incorporating the network of detectors.
Furthermore, in the work\cite{Gao2023}, the fact that the data collection is gradually incremental in real observations is considered. The search results obtained in ahead short period observations can be used to reduce the parameter space in following searches, which can help enhance the efficiency of the algorithm.

Although the framework GBSIEVER has been demonstrated to be capable of extracting a large population of overlapping GCB signals, there are various aspects which can be further improved.
In the following, other two implementations of iterative subtraction focusing on reliving inaccurate contamination and enhancing search efficiency will be introduced.

\subsubsection{Local maxima particle swarm optimization algorithm} \label{subsec_LMPSO}
In the iterative subtraction scheme, the inevitable errors of parameter estimation in each iteration will leave residual signals that can continuously accumulate and contaminate remaining data as noise.
The work \cite{Gao2024} improves the previous GBSIEVER framework and develops the new approach of local maxima particle swarm optimization algorithm with a special strategy of creating voids referred to as LMPSO-CV for dealing with inaccurate subtraction contamination, especially for the low SNR sources.
LMPSO-CV starts with the remaining data assuming all sources with $\text{SNR}>15$ have been identified and subtructed. The algorithm aims at identifying all local maxima of $\mathcal{F}$-statistic in the parameter space, and the source parameters will be extracted from these local maxima.

In LMPSO-CV, to identify the local maxima, the configuration of PSO is adjusted by setting $c_1 = \omega = 0$.
The equations guiding the movement of particles turn into the form of \cite{Gao2024}
\begin{equation} 
    \begin{aligned}
        \boldsymbol{v}_i^{t+1} &= c_2 \boldsymbol{r}_2 \left( \boldsymbol{G}^t - \boldsymbol{x}_i^t\right), \\
        \boldsymbol{x}_i^{t+1} &= \boldsymbol{x}_i^{t} + \boldsymbol{v}_i^{t+1}.
    \end{aligned}
\end{equation}
As mentioned in Section \ref{subsec_iterative}, a lower inertia weight favors the exploitation where particles tend to move within a local promising region, and emphasizing the global term will guide particles to move toward the same position. It can be expected that the setting of entirely dropping the inertia weight and the individual term will extremely improve the convergence towards local maxima.

Once a local maximum is identified, {to avoid the search algorithm picking the same or too close local maxima multiple times, a void in the parameter space will be created and excluded in subsequent searches.}
The void is modeled by a spheroid as
\begin{equation}
    \frac{x^2}{a^2} + \frac{y^2}{b^2} + \frac{z^2}{c^2} = 1, 
\end{equation}
where $x$, $y$, $z$ correspond to the source parameter $f_0$, $\beta$, $\lambda$.
The size of the spheroid is determined by the frequency resolution and the degeneracy of $\beta$ and $\lambda$ through experimental computations.
In the process of searching for local maxima, if a particle moves into voids, it will be placed at a new random position outside voids in the next iteration to avoid redundancy.

\begin{figure*}
    \centering
    \includegraphics[width=\columnwidth]{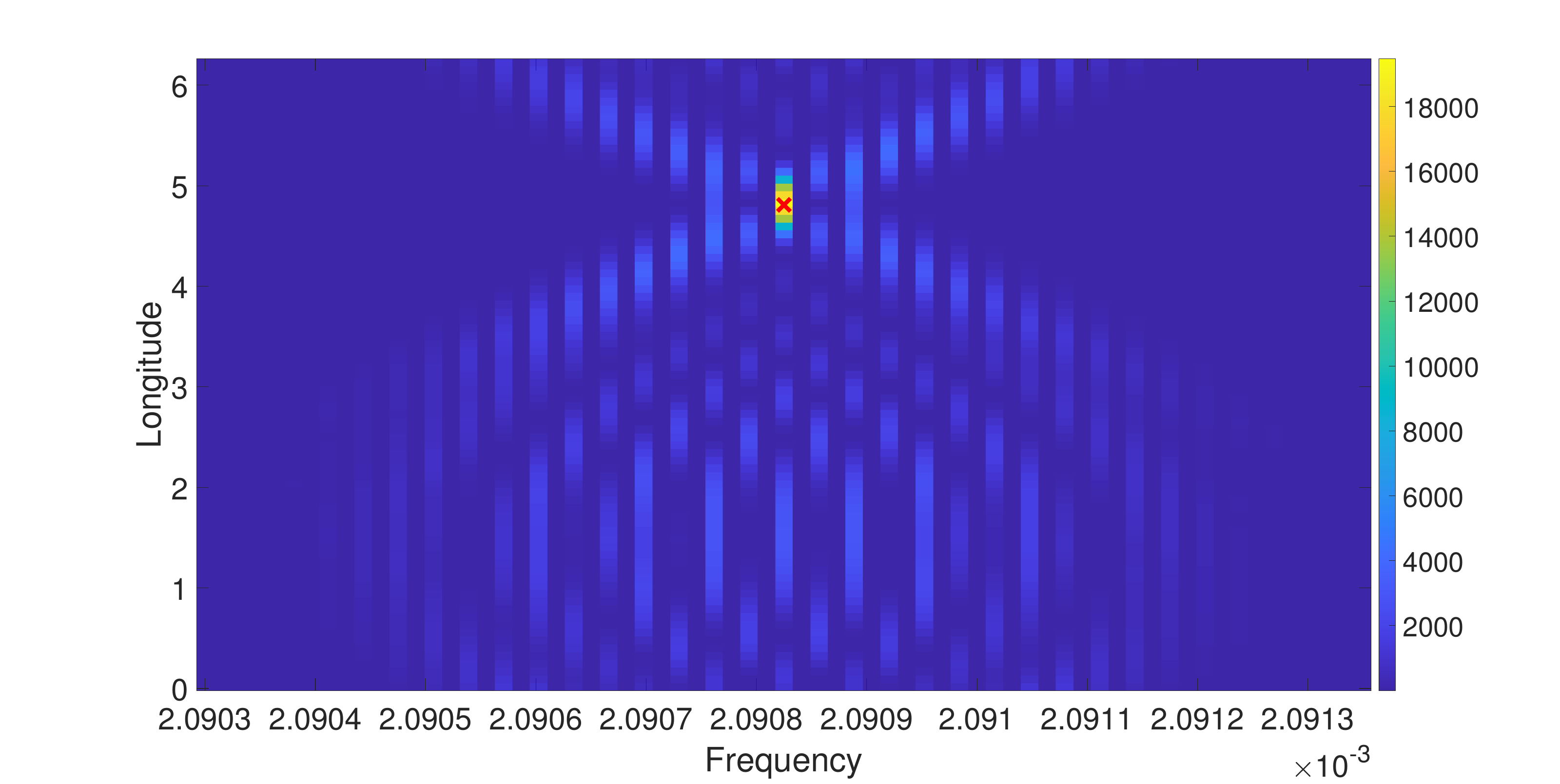}
    \includegraphics[width=\columnwidth]{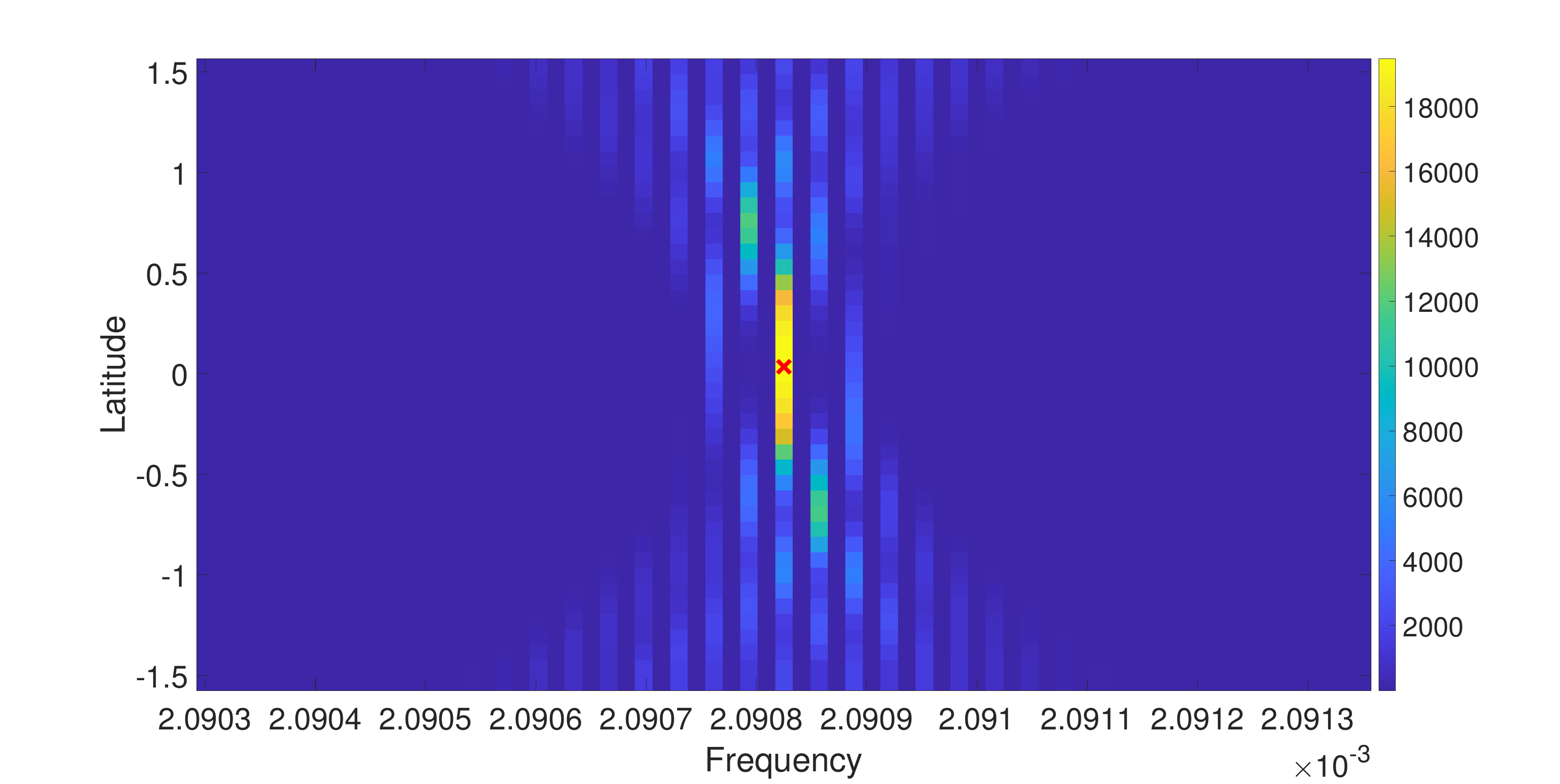}
    \includegraphics[width=\columnwidth]{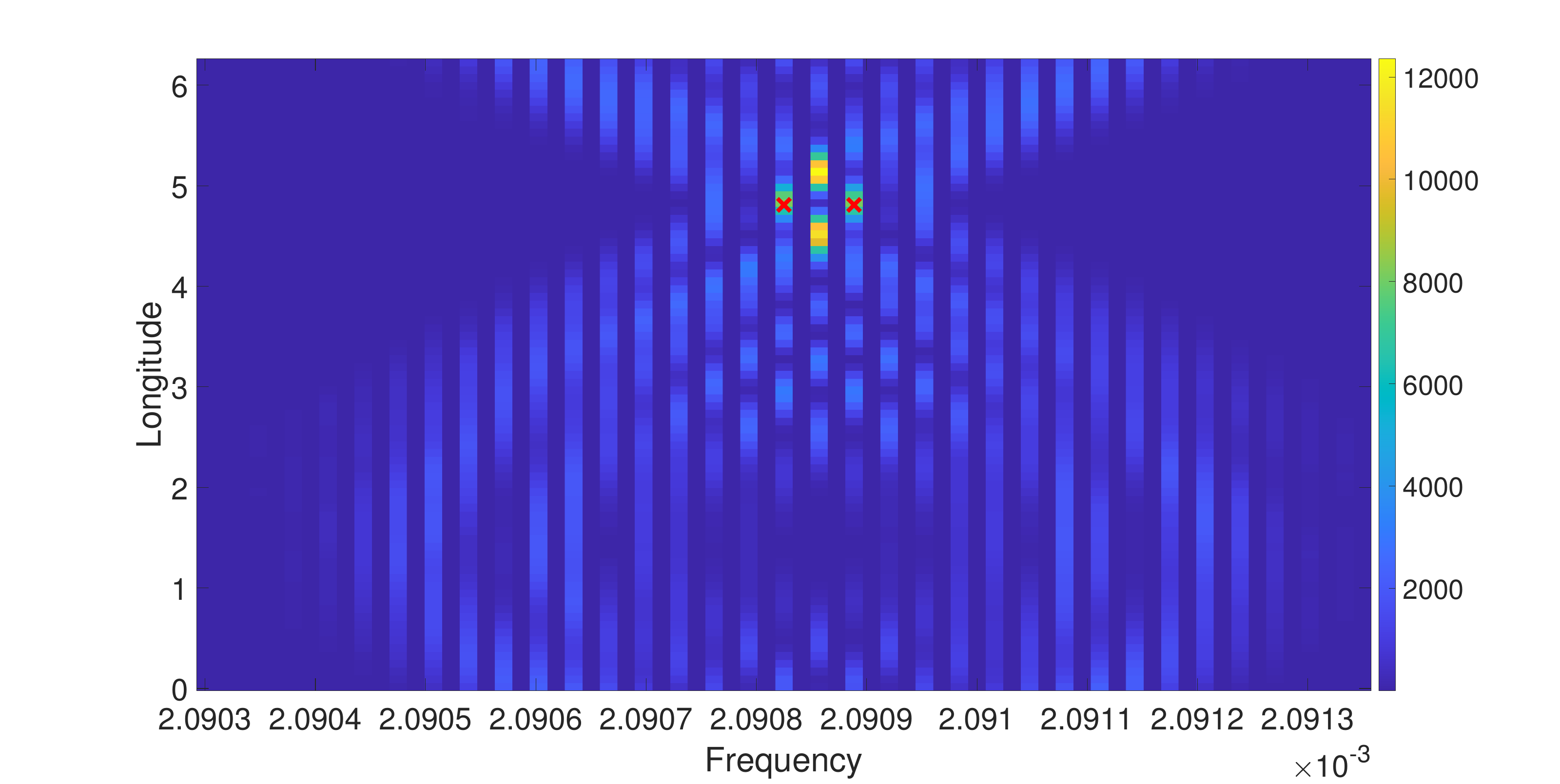}
    \includegraphics[width=\columnwidth]{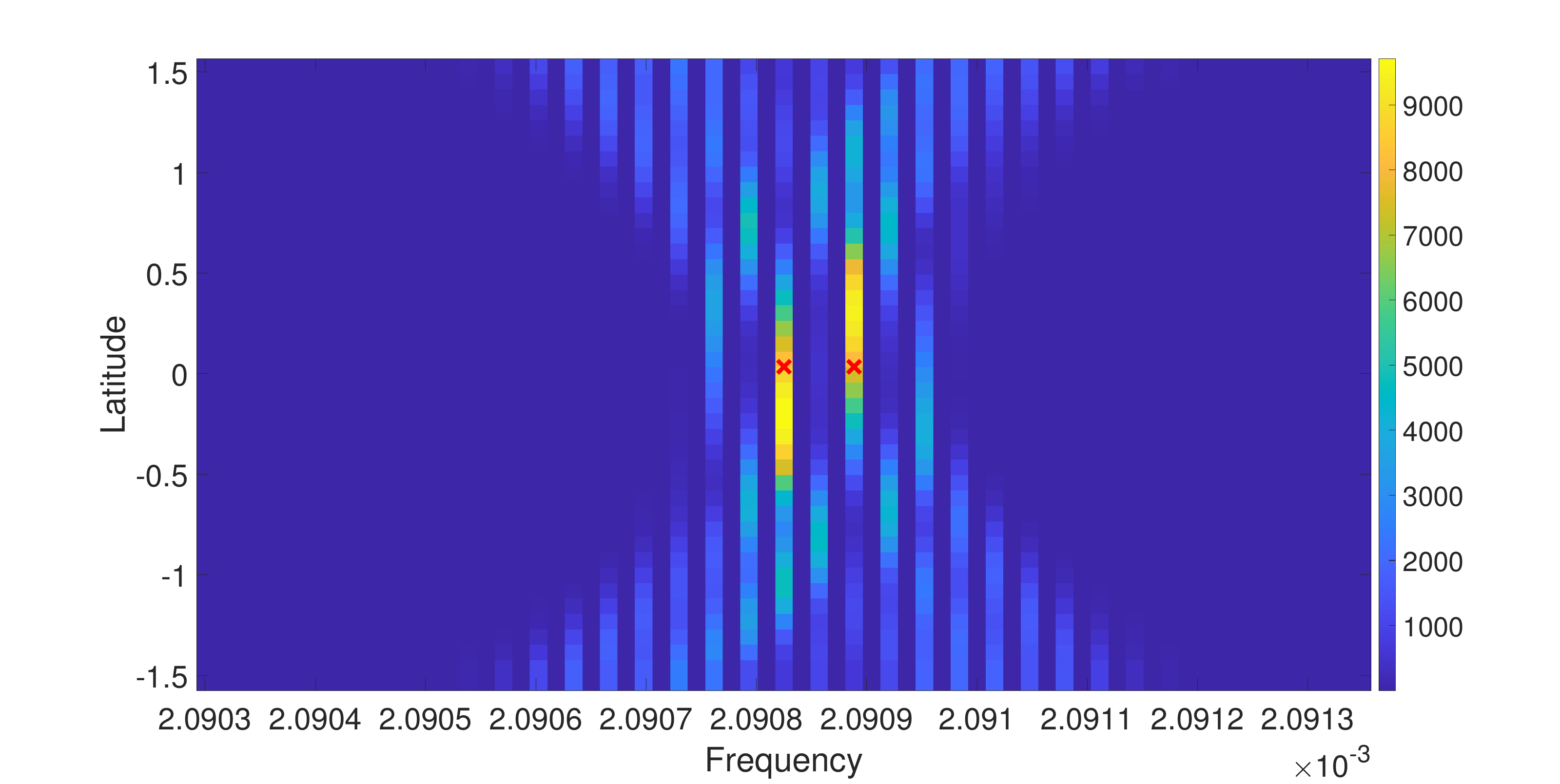}
    \includegraphics[width=\columnwidth]{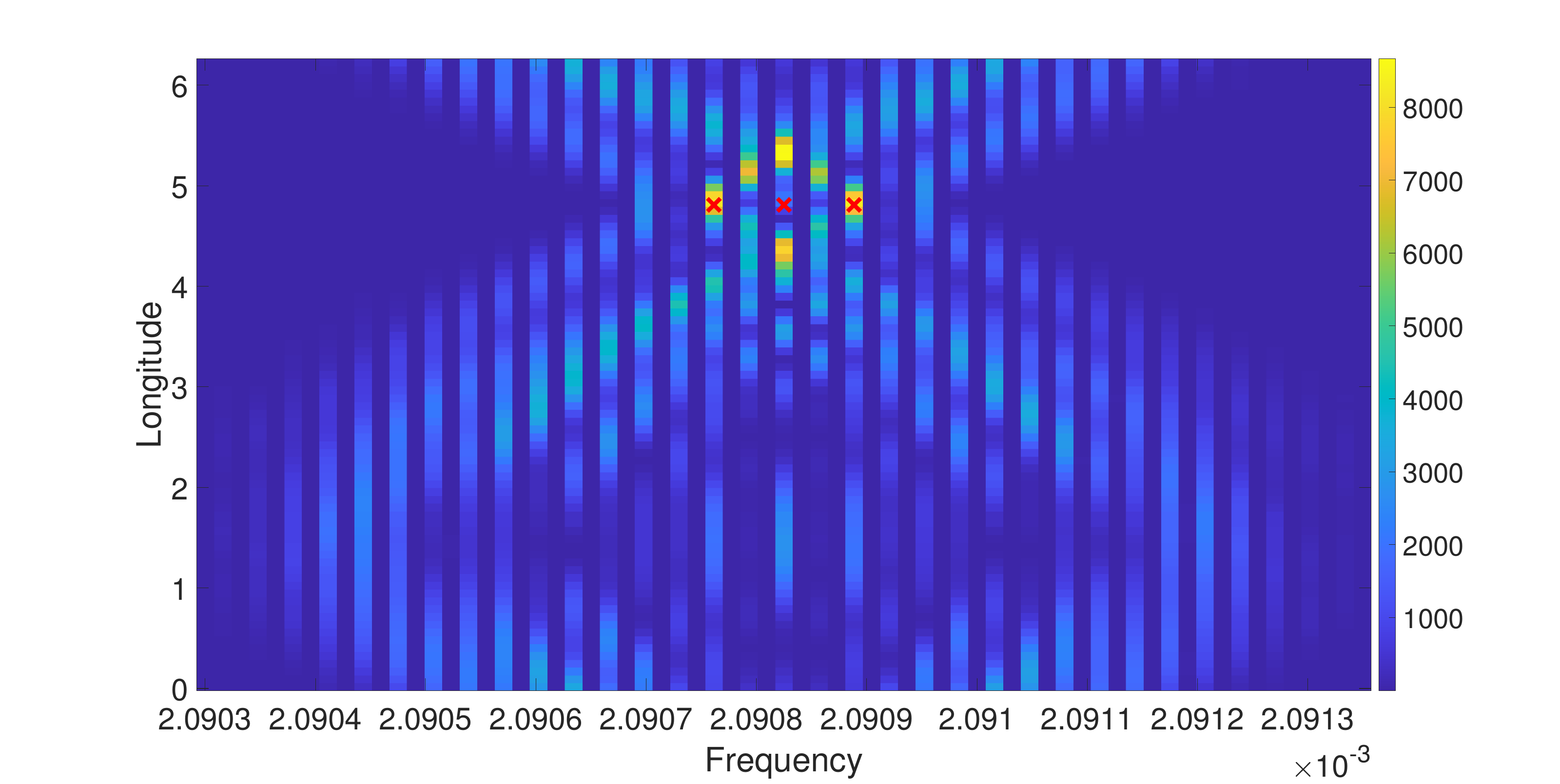}
    \includegraphics[width=\columnwidth]{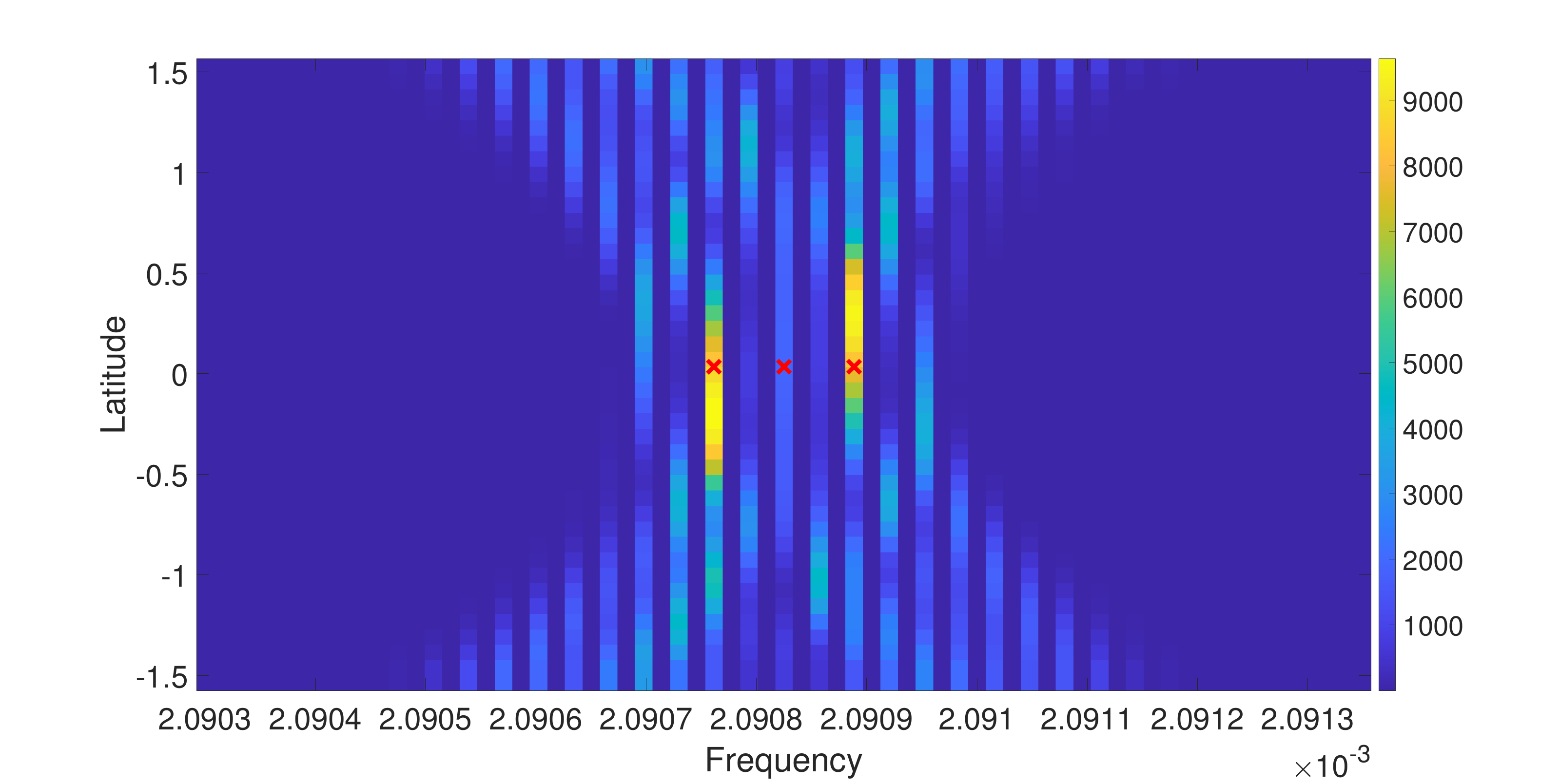}
    \caption{\textbf{The likelihood surfaces for cases of an individual signal and overlapping signals.}
    These figures show the value of the $\mathcal{F}$-statistic likelihood in the parameter space of $(f_0, \lambda, \beta)$. The red crosses denote the injected values, and the colorbars indicate the $\mathcal{F}$-statistic values. The top row is for the case of an individual signal, the middle and bottom rows are for the cases of two and three overlapping signals. 
    For the overlapping signals with enough small frequency intervals, there will be local maxima in likelihood surfaces corresponding to false candidates induced by correlations among overlapping signals, which may have higher likelihood values than the real candidates.
    These figures are cited from \cite{Gao2024}.}
    \label{fig_LMPSO_Fstas_surface}
\end{figure*}

The search of local maxima and creation of voids will be performed repeatedly until the termination rules designed for identifying all local maxima beyond the threshold are satisfied.
The source parameters are extracted among the identified voids.
As can be seen from Figure \ref{fig_LMPSO_Fstas_surface}, there are enormous local maxima in likelihood surfaces corresponding to false signals which need to be eliminated when extracting real candidates.
The extraction consists of three parts.
The first part aims at removing local maxima induced by degeneracy noise of individual signals. 
From the observation of experimental evaluation of the likelihood for an individual signal as shown in the top row of Figure \ref{fig_LMPSO_Fstas_surface}, the local maxima corresponding to false signals have smaller likelihood values than the real signal. 
Therefore, the local maxima are sorted in descending order, and the extraction starts with the local maxima of the highest $\mathcal{F}$-statistic value which is assumed to be a real signal.
Then, the quantities of $\Delta \mathcal{F}$ defined as \cite{Gao2024}
\begin{equation}
    \Delta \mathcal{F}(\boldsymbol{\theta}_i) = \mathcal{F}(\boldsymbol{\theta}_i,\, d(t)) - \mathcal{F}\left(\boldsymbol{\theta}_i,\,\sum_{m=1}^{i-1} h(t, \boldsymbol{\theta_m})\right)
\end{equation}
are computed for subsequent local maxima. 
In the above equation, the first term $\mathcal{F}(\boldsymbol{\theta}_i,\, d(t))$ is the $\mathcal{F}$-statistic value of the $i$-th local maximum with parameter $\boldsymbol{\theta_i}$ with respect to the raw data, which is same to the $\mathcal{F}$-statistic value used in the search process. The second term $\mathcal{F}\left(\boldsymbol{\theta}_i,\,\sum_{m=1}^{i-1} h(t, \boldsymbol{\theta_m})\right)$ is the $\mathcal{F}$-statistic value of the $i$-th local maximum with respect to the summation of candidate signals previous extracted, which characterizes the degeneracy with previous extracted signals.
If the $\mathcal{F}$-statistic value of the $i$-th local maximum is mainly contributed by the degeneracy with previous extracted signals. 
The quantity of $\Delta \mathcal{F}$ is expected to be small.
Conversely, large $\Delta \mathcal{F}$ indicates the local maximum is likely to be a real signal.
The local maxima that satisfy the criterion of $\Delta \mathcal{F}>53$ will be retained and passed into the second part for further removing false signals.

The second part of signal extraction and false signal elimination is based on the astrophysical properties of GCBs.
The second part requires a new independent search of local maxima which is similar to the process discussed above except using a different range for $\dot{f}$.
In the first search, the search range for $\dot{f}$ is $[-10^{-16}, 10^{-15}]$ when $f\le4$ mHz and $[-10^{-14}, 10^{-13}]$ when $f>4$ mHz. Whereas, in the second search, the range of $\dot{f}$ is determined by the mass parameters of binaries.
Since the majority of GCBs are consisted of binary white dwarfs which have the mass ranging from $0.1$ to $1.4M_\odot$. In the second search, the values of $\dot{f}$ are constrained within the range corresponding to the mass range of white dwarfs.
After two independent searches, the candidate signals need to be compared through the correlation coefficient defined as 
\begin{equation}
    R(\boldsymbol{\theta}, \boldsymbol{\theta}') = \frac{C(\boldsymbol{\theta}, \boldsymbol{\theta}')}{[C(\boldsymbol{\theta}, \boldsymbol{\theta})C(\boldsymbol{\theta}', \boldsymbol{\theta}')]^{\frac12}},
\end{equation}
where $C(\boldsymbol{\theta}, \boldsymbol{\theta}') = \langle h(\boldsymbol{\theta}, \, h(\boldsymbol{\theta}')\rangle$.
This correlation coefficient quantifies how similarity between two signals identified in two independent searches.
Only candidates with enough large $R$ are considered as reliable sources and are retained for further processing.
Another characterization of GCBs is the concentration in the Galactic disk.
Therefore, one can simply remove candidates with Galactic latitude outside of $[-0.5\text{rad}, 0.5\text{rad}]$ to enhance the reliability.

The third part of signal extraction focuses on the false candidates induced by overlapping signal degeneracy noise.
In the experimental computation of the likelihood surface for the situation of overlapping signals as shown in the middle and bottom rows of Figure \ref{fig_LMPSO_Fstas_surface}, it is found that the $\mathcal{F}$-statistic values of local maxima corresponding to false candidates can be even larger than real signals. To remove false candidates of this kind, a similar process of the first part needs to be performed again but in ascending order of local maxima.

The process of LMPSO-CV can be summarized in two steps.
In the first step, the stochastic optimization algorithm is tuned to specializing in finding local maxima. A void in parameter space will be created to avoid redundant search once a local maximum is identified.
The second step extracts signals and eliminates false candidates from the identified local maxima, which consists of three parts: removing false candidates induced by degeneracy of individual signals, applying the astrophysical properties of GCBs, and removing false candidates induced by degeneracy of overlapping signals.
LMPSO-CV mainly focuses on the sources with SNR between 7 and 15.
Since the LMPSO-CV approach searches all local maxima simultaneously and filters out false candidates afterward, rather than iteratively performing search and subtraction for a single signal, it can better deal with inaccurate subtraction contamination in the traditional iteration subtraction scheme.

\subsubsection{Combination of coarse template search and fine PSO search} \label{subsec_coarse_fine}
Another implementation of the iterative subtraction scheme focuses on improving search efficiency by the strategy of combination with the coarse search based on templates for quickly identifying rough parameters of candidates and the fine search based on $\mathcal{F}$-statistic and PSO for detailed exploring small promising regions \cite{Lu2022}.

The strategy is composed of two steps.
In the first step, candidates are searched by a template bank.
A stochastic template bank \cite{Messenger2009} is constructed, where the places of template parameters are randomly drawn in the parameter space and additional pruning operation is performed to remove templates with too small separation.
The number of templates can be approximated by \cite{Lu2022}
\begin{equation}
    N(\eta, m_*, \mathcal{S}_n) \approx \frac{V_{\mathcal{S}_n}}{V_n}\ln \left(\frac{1}{1-\eta}\right) m_*^{-n/2},
\end{equation}
where $n$ denotes the dimension of the parameter space $\mathcal{S}_n$, $\eta$ denotes the desired level of coverage confidence, $m_*$ is the mismatch criterion, $V_n$ is the volume of unit sphere in $n$-dimension, and $V_{\mathcal{S}_n}$ is the volume of the whole parameter space.
The construction of the template bank starts with randomly generating $N$ waveforms.
Then, {the KDtree (k-dimensional tree) algorithm \cite{Maneewongvatana1999, Bentley1975} is used to find the nearest neighbor for each waveform.} If the mismatch between two waveforms is less than $m_*$, one of them will be removed.
The above two procedures are performed repeatedly until the total number of templates gets to stable.
The $\mathcal{F}$-statistic likelihood is also used in the coarse search, the threshold for coarse search is set to $\mathcal{F}_{\text{th\_coarse}} =15$ which corresponds to $\text{SNR}\approx 5.1$. The templates whose $\mathcal{F}$-statistic values exceed the threshold are recorded as candidates.
Due to the annual orbital motion of detectors, modulation is imposed on the quasi-monochromatic signals of GCBs.
The modulation can induce sidebands around the central frequency $f_0$ of GCB signals.
A clustering operation will be performed to gather candidates with close central frequency.
Among all candidates within the extended Doppler window, only the sources with the largest $\mathcal{F}$-statistic values are retained for fine search in the next step.

In the second step, the parameters of candidates identified by the template search are accurately estimated using PSO, which is similar to the procedure discussed before, while the search ranges for the central frequency are tuned according to results given by coarse search.

The performance of this methodology is demonstrated with MLDC3.1. It is shown that $\mathcal{O}(10^4)$ sources can be successfully identified, and nearly 90 percent of them are well aligned with injected signals.
This method performs a coarse template search in the first step, which can provide priori information guiding the PSO search thus enhancing the efficiency of the fine parameter estimation.

\subsection{Global fitting with trans-dimensional Bayesian inference} \label{subsec_global}
The iterative subtraction method has advantages of rapidity and efficiency but also suffers from problems of correlations among overlapping signals and contamination of inaccurate subtraction.
In each step of iterative subtraction, errors in parameter estimation are unavoidable, which can yield signal residuals left behind the subtraction. The residuals will contaminate remaining data and accumulate along with iterations. 
Furthermore, in each iteration only the parameter values with maximum likelihood are used in subtraction, the uncertainty of parameter fitting will propagate alone iterations. The uncertainty of parameter estimation for identified candidates is difficult to ascertain. 
Additionally, GCB signals are heavily overlapping in both time and frequency, which can induce high correlations among signals.
The correlations may lead to bias in the parameter estimation of each iteration, and this will further intensify the problem of inaccurate subtraction contamination.

To alleviate these problems, a commonly used operation is dividing the whole frequency band into small bins and performing analyses independently in different bins while carefully addressing signals residing at edges, which can reduce the number of iterations and reduce the accumulation of inaccurate subtraction contamination.
Besides, as introduced in Section \ref{subsec_LMPSO}, the method of modifying the particle movement rule in PSO aiming at identifying all local maxima on the likelihood surface simultaneously is also an effective method to address inaccurate subtraction contamination for the low SNR sources.

The global fitting is another route that can effectively deal with the problems of overlapping signal correlations.
In the iterative subtraction scheme, only parameters for one source are estimated in each step. In contrast, parameters for all sources together with the source number are estimated simultaneously in the global fitting scheme.
The global fitting with full Bayesian parameter estimation can obtain joint posterior distributions rather than just the maximum likelihood estimation, thus the uncertainty of fitting can be read out from posteriors straightforwardly. Since all sources are fitted simultaneously, the correlations of overlapping signals are taken into account and can be reflected in joint posteriors of multiple overlapping signals.

As mentioned in \ref{subsec_early}, the strategy of global fitting and the Bayesian approach have been proposed and implemented early in various works focusing on MLDCs \cite{Umstaetter2005a,Umstaetter2005,Cornish2005,Crowder2006,Crowder2007,Littenberg2011}.
Further improvements like incorporating the reality that the data collection is time evolving, various sophisticated methods for increasing the efficiency of MCMC sampling, and new implementations of RJMCMC have been continuously presented in recent years \cite{Littenberg2020,Littenberg2023,Lackeos2023,Karnesis2023,Katz2024}.

\subsubsection{RJMCMC}
For the Bayesian inference of global fitting, One of the key differences with the Bayesian parameter estimation commonly used in data analyses of current ground-based detectors is that the number of sources is uncertain and has to be inferred from the given data.
A single GW source signal $h(t, \boldsymbol{\theta})$ in the likelihood shown in Equation \ref{eq_likelihood} turns to a summation of multiple signals $\sum_k h(t, \boldsymbol{\theta}_k)$ where the number of source is uncertain.
The sampling algorithms to solve inference problems with uncertain dimensionality are referred to as trans-dimensional MCMC or RJMCMC.

In the parameter estimation problem of the Bayesian framework, the estimation results are represented by posteriors through the Bayes theorem
\begin{equation}
p(\boldsymbol{\theta}_n|d(t)) = \frac{\pi(\boldsymbol{\theta}_n) p(d(t)|\boldsymbol{\theta}_n)}{\mathcal{Z}},
\end{equation}
where $d(t)$ denotes observed data, $\boldsymbol{\theta}_n$ represents parameters with the dimensionality of $n$, $\pi(\boldsymbol{\theta}_n)$ is the prior representing the knowledge before observations, $p(d(t)|\boldsymbol{\theta}_n)$ is the likelihood representing the probability of noise realization that can just present the time series $d(t)$ observed on detectors when adding the GW signals described by parameters $\boldsymbol{\theta}_n$ and depending on the models of noise behavior and GW signals, $\mathcal{Z}$ is the normalization factor called as evidence.
Since the dimension of parameters $\boldsymbol{\theta}_n$ for describing GW signals is usually high, it is impractical to compute posterior on a grid of the parameter space. Stochastic sampling algorithms are usually employed to perform random walking in the parameter space, and use the density distributions of random samples to approximate the probability distributions of posteriors.

The Metropolis–Hasting MCMC (MHMCMC) \cite{Hastings1970,Metropolis1953} is one of the well known and widely used algorithms to perform sampling for fixed dimensional problems.
After setting the initial state of a random walker, the subsequent movements are guided by the rules introduced below.
If current state is $\boldsymbol{\theta}^{i}_n$, a new state can be drawn from a proposal distribution $q(\boldsymbol{\theta}^{i+1}_n|\boldsymbol{\theta}^{i}_n)$.
Then the acceptance probability of the new state is evaluated by
\begin{equation}
    \alpha = \min\left[1, \frac{p(\boldsymbol{\theta}^{i+1}_n|d) \, q(\boldsymbol{\theta}^{i}_n|\boldsymbol{\theta}^{i+1}_n)}
    {p(\boldsymbol{\theta}^{i}_n|d) \, q(\boldsymbol{\theta}^{i+1}_n|\boldsymbol{\theta}^{i}_n)} \right].
\end{equation}
The probability of whether actually moving from current state $\boldsymbol{\theta}^{i}_n$ to new proposed state $\boldsymbol{\theta}^{i+1}_n$ is determined by $\alpha$.
It can be proven that after sufficient walking, the equilibrium distribution of the walker will converge to the target posterior distribution $p(\boldsymbol{\theta}_n|d)$.
Although, in practical problems, the MHMCMC usually needs to be modified in various aspects to increase sampling efficiency or avoid bias, the above description presents the most basic conception of the MCMC sampling algorithm for fixed dimensional problems.

In problems of uncertain dimensionality, to allow the walker to jump between different parameter spaces, the RJMCMC algorithm draws new proposals by a different procedure \cite{GREEN1995,Sambridge2006}.
Firstly, a random vector $\boldsymbol{u}$ with the dimension of $r$ is generated from a chosen probability distribution $g(\boldsymbol{u})$. 
Then, the proposal state is generated through a transformation
\begin{equation}
    \boldsymbol{\theta}^{i+1}_{n'} = f(\boldsymbol{\theta}^{i}_{n}, \boldsymbol{u}).
\end{equation}
The corresponding inverse transformation is given by
\begin{equation}
    \boldsymbol{\theta}^{i}_{n} = f^{-1}(\boldsymbol{\theta}^{i+1}_{n'}, \boldsymbol{u}'),
\end{equation}
where $\boldsymbol{u}'$ is a vector with size $r'$ and satisfying $n+r=n'+r'$. 
The only requirement for the transformation is that $f$ and $f^{-1}$ need to be differentiable.
The dimensions of parameter space $n$ and $n'$ do not need to be the same, and even the parameterization associated with  $\boldsymbol{\theta}^{i}_{n}$ and $\boldsymbol{\theta}^{i+1}_{n'}$ can be different. 
The acceptance ratio for the new proposed state is given by
\begin{equation}
    \alpha = \min\left[1, \frac{p(\boldsymbol{\theta}^{i+1}_{n'}|d) \, g'(\boldsymbol{u}')}
    {p(\boldsymbol{\theta}^{i}_{n}|d) \, g(\boldsymbol{u})} \left|\boldsymbol{J}\right| \right].
\end{equation}
The term $\boldsymbol{J}$ is the Jacobian defined by
\begin{equation}
    \boldsymbol{J} = \left| \frac{\partial (\boldsymbol{\theta}^{i+1}_{n'}, \, \boldsymbol{u}')}{\partial (\boldsymbol{\theta}^{i}_{n}, \, \boldsymbol{u})}\right|
\end{equation}
to account for the change of volume element under the transformation of parameter spaces.
In general, the Jacobian can be difficult to compute. 
However, in the situation of global fitting for GCB signals, where it only involves adding or removing GCB signals with the same set of parameters, the Jacobian can be easily computed by the ratio of prior volumes of parameter spaces before and after the transformation.
The walker of RJMCMC can freely explore the entire possible parameter space including not only parameter spaces of individual sources but also parameter spaces for all possible source numbers.
The estimation of source number can be naturally obtained through comparing the iteration numbers of the walker lingered in the different parameter spaces, and the Bayesian factor for the assumptions of different source numbers can be given by the ratio of the iteration numbers in different parameter spaces.

Although the movement rule of RJMCMC can be explicitly presented as above, the implementation of the RJMCMC algorithm in practice to solve the problem of overlapping GCB signals is very challenging. 
On the one hand, since all sources are simultaneously fitted, the parameter spaces are extremely high dimensional. Even the whole frequency band is usually divided into small bins and different bins can be analyzed independently after carefully addressing edge effects. The source number in a single small bin can still be considerable. For example,  the default setting of the prior range for the source number in each bin is $[0, 30]$ in the work \cite{Littenberg2020}. As mentioned in Section \ref{sec_GCBs}, a GCB signal is characterized by 8 parameters. The maximal dimension of parameter space can be 240. 
And the posterior can have the feature of multimodality, which requires the sampler has the ability to deal with complicated likelihood surfaces in high-dimensional parameter spaces. 
On the other hand, the uncertainty of source number requires the sampler to jump between different parameter spaces, which further extremely expands the space needed to explore.
Furthermore, since the likelihood in vast regions may be very small.
Only if the proposal is enough close to the true values, it can likely be accepted.
Generating proposals entirely uniformly may lead extremely low acceptance rate, especially for between-model proposals.
One need to design good strategies for random walking to increase the sampling efficiency.

\subsubsection{Global fitting piplines}
Currently, available full-scale and end-to-end global fitting pipelines for GCBs include GBMCMC \cite{Littenberg2020,Littenberg2023,Lackeos2023} and Eryn \cite{Karnesis2023,Katz2024}.
Various intelligence methods are utilized to overcome difficulties in the global fitting with RJMCMC.
For example, the method of parallel tempering \cite{Vousden2015,Hukushima1996,Swendsen1986} is used in both GBMCMC and Eryn.
In parallel tempering, multiple walkers are randomly moving in parallel at different temperatures $T$.
The likelihood is modified as
\begin{equation}
    p_T(\boldsymbol{\theta}|d) \propto \pi(\boldsymbol{\theta}) p(d|\boldsymbol{\theta})^{1/T},
\end{equation}
{where $\pi(\boldsymbol{\theta})$ is the prior and $p(d|\boldsymbol{\theta})$ is the original likelihood function.}
When $T=1$, $p_T(\boldsymbol{\theta}|d)$ returns to the target posterior distribution.
When $T \rightarrow \infty$, $p_T(\boldsymbol{\theta}|d)$ approaches to the prior distribution. 
Higher temperatures correspond to more exploration where walkers can escape from local optima more easily, while lower temperatures correspond to more exploitation where walkers can sample the small promising regions in more detail.
During the random walking, state exchanges are attempted between walkers at different temperatures. 
The probability of acceptance of exchange proposals is determined by the ratio
\begin{equation}
    \alpha = \min \left[ 1, \, \frac{p_{T_i}(\boldsymbol{\theta}_i) p_{T_j}(\boldsymbol{\theta}_j)} 
    {p_{T_i}(\boldsymbol{\theta}_j) p_{T_j}(\boldsymbol{\theta}_i)}\right].
\end{equation}
The detailed balance is maintained under this state exchange between walkers, which ensures that the equilibrium distribution of walkers can converge to the target posterior distributions.
The temperature ladder needs to be chosen for maximizing the information flow among walkers at different temperatures.
Ideally, one may expect an equal acceptance ratio between every pair of walkers with neighboring temperatures.
The GBMCMC and Eryn adopt the scheme presented in \cite{Vousden2015} to set the temperature ladder, where the temperature space is dynamically adjusted in the initial burn-in phase to obtain a stable configuration for the rest of random walking.
The parallel tempering mechanism can help samplers efficiently explore the complicated likelihood surface with high multimodality.

The strategies for drawing proposals play a crucial role in the sampling efficiency.
Customized proposal distributions are developed in GBMCMC to enhance the sampling efficiency.
Ideal proposal distributions would be identical to posterior distributions or likelihood functions.
The likelihood of multiple overlapping signals consists of three parts, the correlation between each signal and noise, the correlation of noise itself, and the cross-correlations among overlapping signals.
As discussed in \cite{Littenberg2020}, high correlations among overlapping signals are relatively rare.
Therefore, one can enhance proposal distributions focusing on individual signals.
Although the cross-correlations of overlapping signals are neglected when designing the proposal distributions, the sampler still explores the joint parameter space of multiple overlapping signals which incorporates cross-correlations of overlapping signals.

\begin{figure*}
    \centering
    \includegraphics[width=2\columnwidth]{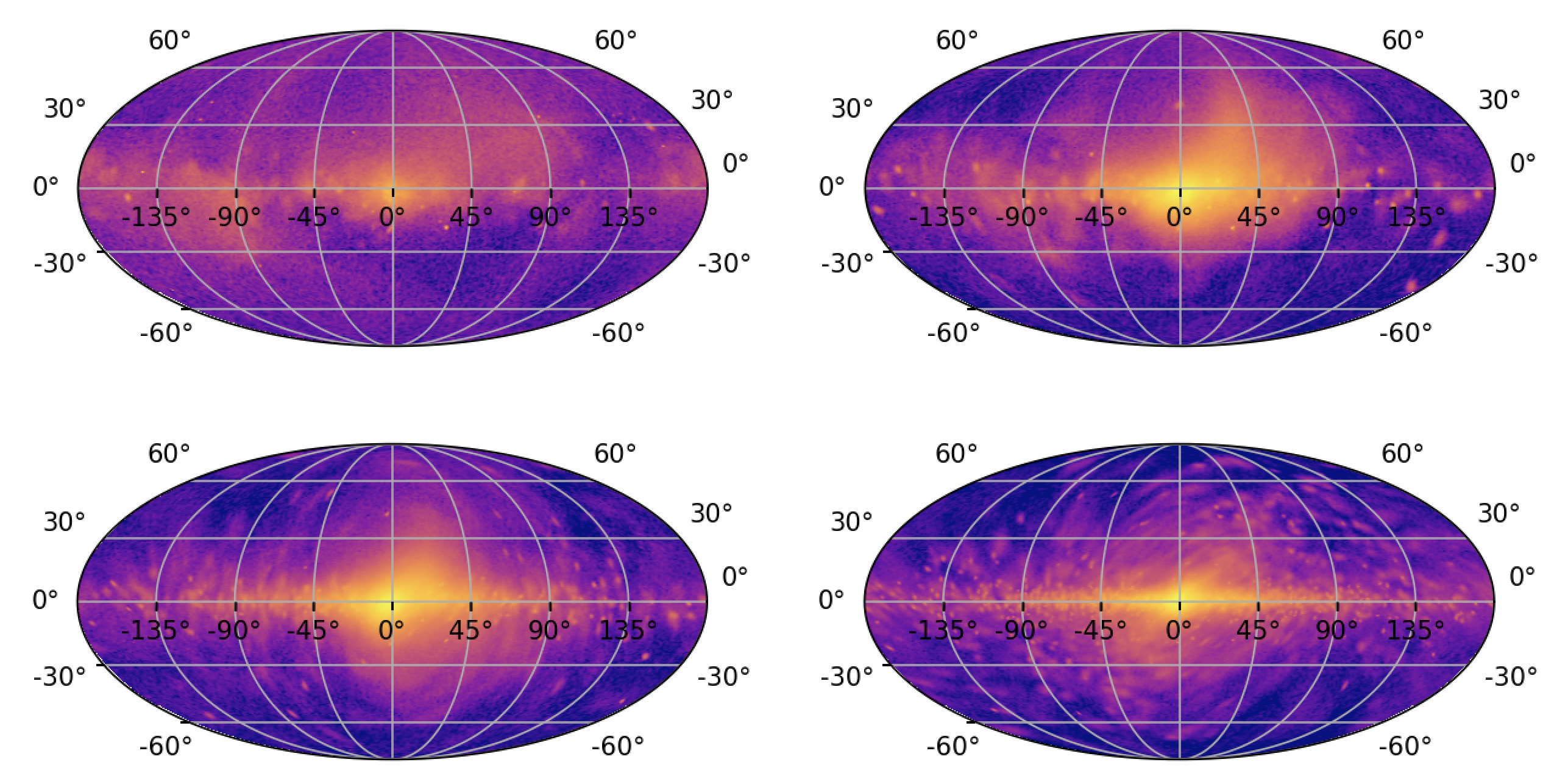}
    \caption{\textbf{Posteriors of sky location of identified GCBs in the Galactic coordinate given by GBMCMC}. 
    The subplots correspond to the observations of 1.5 (top left), 3 (top right), 6 (bottom left), and 12 (bottom right) months.
    These figures are cited from \cite{Lackeos2023}}.
    \label{fig_GBMCMC_sky}
\end{figure*}

The design of proposal distribution in GBMCMC utilizes the $\mathcal{F}$-statistic, the feature of multimodality due to orbital motion of detectors, and posteriors from former epochs of observation.
The method of $\mathcal{F}$-statistic is widely used in various implementations of iterative subtraction as presented in previous sections, it can also be used here to construct the proposal distribution for a single source.
GBMCMC uses the $\mathcal{F}$-statistic to build proposals for parameters ($f_0$, $\lambda$, $\beta$) which are precomputed on a grid with spaces determined by estimation of the Fisher matrix.
The values of $\mathcal{F}$-statistic can be approximated to the original likelihood values, thus the $\mathcal{F}$-statistic proposals are expected to have a high accepted rate.
As mentioned in Section \ref{subsec_coarse_fine}, the modulation induced by orbital motion can cause sidebands around the central frequency $f_0$ of quasi-monochromatic GCB signals.
The likelihood surface have the feature of multiple modes around $f_0$.
GBMCMC designs a dedicated proposal to update the frequency parameter by shifting $f_0$ with modulation frequency $f_\text{m} $ as $f_0 \rightarrow f_\text{0} +nf_\text{m}$, where $f_\text{m} = 1/\text{year}$.
The multimodal proposal can address this known degeneracy and help to improve the convergence of the sampler.
In reality, the data collection is gradually incremental. It is impractical to wait until after the finish of entire observation to analyze the obtained data.
Data analyses must be performed accompanying with data accumulation.
Therefore, one can utilize posteriors obtained from former periods of observation to construct posterior-based proposal distributions in parameter estimation with more observed data, which can significantly improve the sampler convergence.
The results of GBMCMC with simulated GCB dataset are demonstrated in Figure \ref{fig_GBMCMC_sky} where the posteriors of sky location of identified GCBs are plotted.
It can be seen that with the incremental data, the GCBs can be better localized and the structure of the Galaxy is clearer indicated.

Another independent implementation of RJMCMC referred to as Eryn \cite{Karnesis2023,Katz2024} which adopts different sophisticated mechanisms in stochastic sampling to overcome poor convergence in RJMCMC.
Eryn is based on the ensemble sampler \texttt{emcee} \cite{ForemanMackey2013} which uses multiple interacting walkers exploring the parameter space simultaneously with the so-called stretch-move proposal to enhance the efficiency and convergence of the sampling.
Whereas, in order to address the trans-dimensional movement and the heavily multimodal likelihood surface of overlapping GCBs, Eryn extends the stretch-move proposal in origin emcee sampler to the group proposal which sets a stationary group of walkers and makes the state updates likely to happen within one same mode in the likelihood surface.
Furthermore, Eryn build efficient proposals through a data-driven approach, where proposals can be drawn from a fitted distribution constructed by posteriors obtained in burn-in runs with residual data after subtracting bright sources. 
To overcome the problem of low acceptance rate, Eryn implements two mechanisms called delayed rejection \cite{Green2001,Trias2009a} and multiple try metropolis \cite{Martino2018,Liu2000,Bedard2012,Martino2012}.
However, due to the high computational requirement of delayed rejection, although this mechanism is implemented in Eryn, it is not currently used in solving the problem of global fitting for GCBs.
Another important feature of Eryn is the utilization of GPU which reduces the wall-time to perform the whole analysis by parallel computing on contemporary computational hardware.

\begin{figure*}
    \centering
    \includegraphics[width=2\columnwidth]{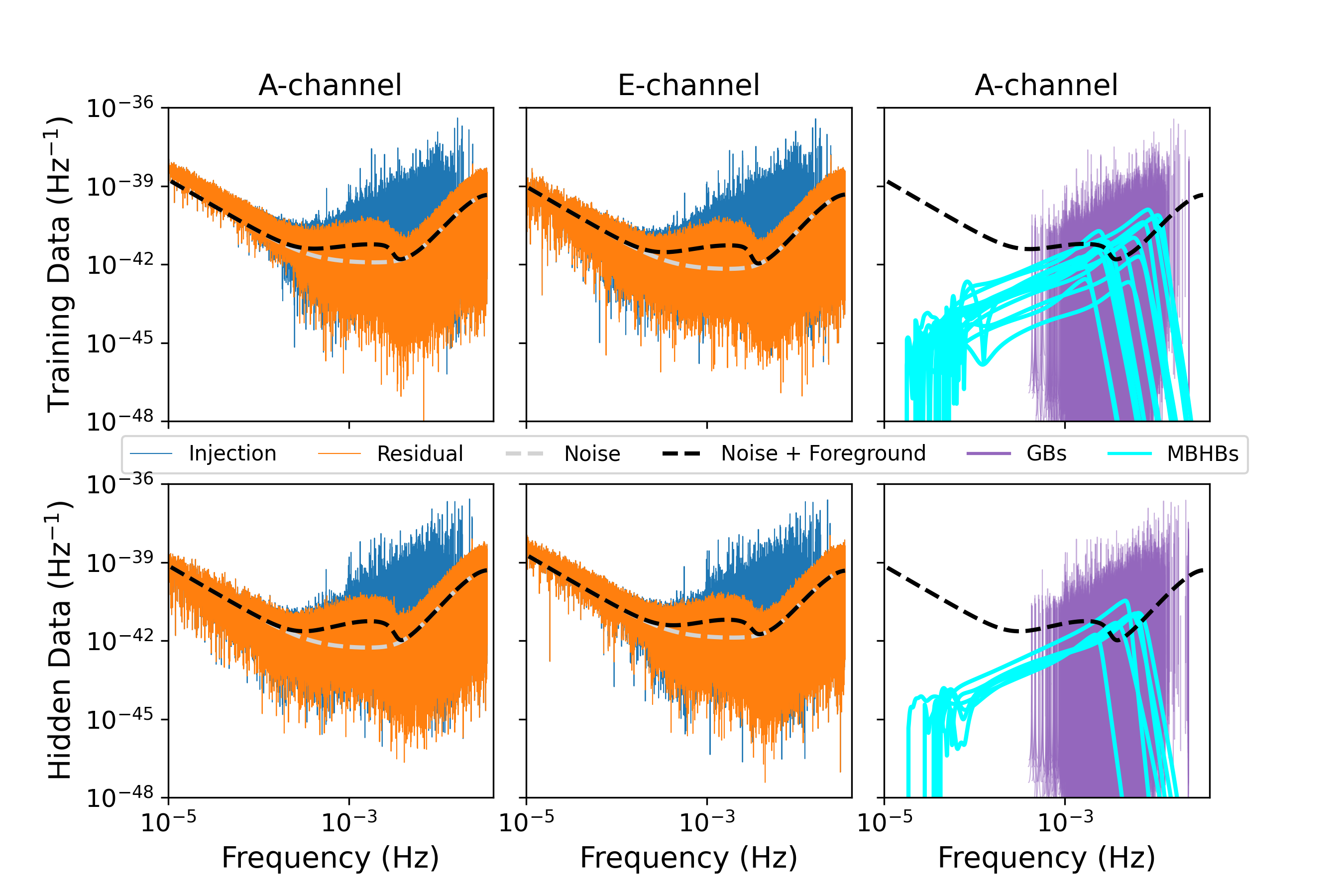}
    \caption{\textbf{Results given by a global fitting pipeline incorporating GCBs and MBHBs.}
    The top and bottom rows show the results of training and hidden datasets in the LDC2a respectively.
    The left and middle columns present the injection data, the residual after subtracting identified signals, and the fitting of noise curves. The identified signals are illustrated in the right column.
    This figure is cited from \cite{Katz2024}.}
    \label{fig_Eryn_residual}
\end{figure*}

The ultimate goal of global fitting is simultaneously analyzing all kinds of sources contained in data including MBHBs, EMRIs, and even unmodeled signals, etc. 
In the work \cite{Littenberg2023} and \cite{Katz2024}, GBMCMC and Eryn are incorporated within global fitting pipelines that can handle blended data of MBHBs and GCBs.
Global fitting for multiple source types can be realized by the blocked Gibbs scheme considering that correlations among different source types are relatively small.
Parameter estimations for different source types can be implemented in separate modules, and these modules are assembled by a wheel update strategy where one can perform updates in one module conditioning on fixed other modules and all modules are updated periodically.
Figure \ref{fig_Eryn_residual} shows the results given by a global fitting pipeline incorporating GCBs and MBHBs where the fitting of GCBs is accomplished by Eryn.

In a short summary, fitting overlapping GCB signals through the full-Bayesian approach can be realized by RJMCMC.
Practical implementations of RJMCMC have to overcome various difficulties such as the heavy multimodality of likelihood surface in vast parameter space, low acceptance rate of trans-dimensional proposals, and poor convergence of random walker, etc. which requires elaborate MCMC algorithms.
Prototype global fitting pipelines for GCBs have been accomplished by GBMCMC and Eryn which can well address simulated data and are successfully incorporated within the global fitting pipelines for blended data of MBHBs and GCBs.
The global fitting with RJMCMC is a full Bayesian approach and can provide joint posteriors of overlapping signals, which can well account for correlations among overlapping GCBs and avoid inaccurate subtraction contamination in the iterative subtraction scheme.
Whereas the full Bayesian approach requires massive computational cost and complicated MCMC algorithms which are difficult to implement.

\subsection{Hybrid Bayesian approach} \label{subsec_hybrid}

The iterative subtraction strategy offers solutions with high efficiency where the computation burden is modest and analysis can be finished in a relatively short time.
Whereas, in each iteration, only the maximum likelihood estimation for a single source is extracted, the correlation of overlapping signals can not be well accounted for. The inaccuracy of estimation in each iteration will contaminate the remaining data. And uncertainty analyses for the identified signals are difficult. 
The full Bayesian approach with RJMCMC where overlapping signals are fitted simultaneously can effectively overcome these difficulties, but with the cost of extremely massive demand on computational resources.
In the works \cite{Strub2022,Strub2023,Strub2024}, a hybrid approach where the maximum likelihood estimation is performed first to find the approximate values of signal parameters, and then the MCMC sampling is used to obtain posteriors of signal parameters is proposed aiming at combining the advantages of the maximum likelihood estimation and the Bayesian parameter estimation while evading their drawbacks.

Similar to the solutions of iterative subtraction introduced previously, the hybrid Bayesian approach also needs to identify the maxima of likelihood in the first step.
However, the methods used for searching maxima are different.
In this hybrid Bayesian approach, the off-the-shelve algorithms of the differential evolution (DE) \cite{Storn1997,Qiang2014} and the sequential least squared programming (SLSP) \cite{Bonnans2006} implemented in the \texttt{scipy} library \cite{Virtanen2020} are adopted for searching the maxima in the likelihood surface. 
The DE method is used to identify the candidates at first through the iterative subtraction scheme, where only one signal is fitted each time the fitting is repeatedly performed with the remaining data after subtracting the best-fit signal. 
Then, in order to address the correlation of overlapping signals, using all found candidates as the start, the global optimizations are performed by the SLSP method to search the maximum likelihood in the joint parameter space of all candidates.

\begin{figure*}
    \centering
    \includegraphics[width=2\columnwidth]{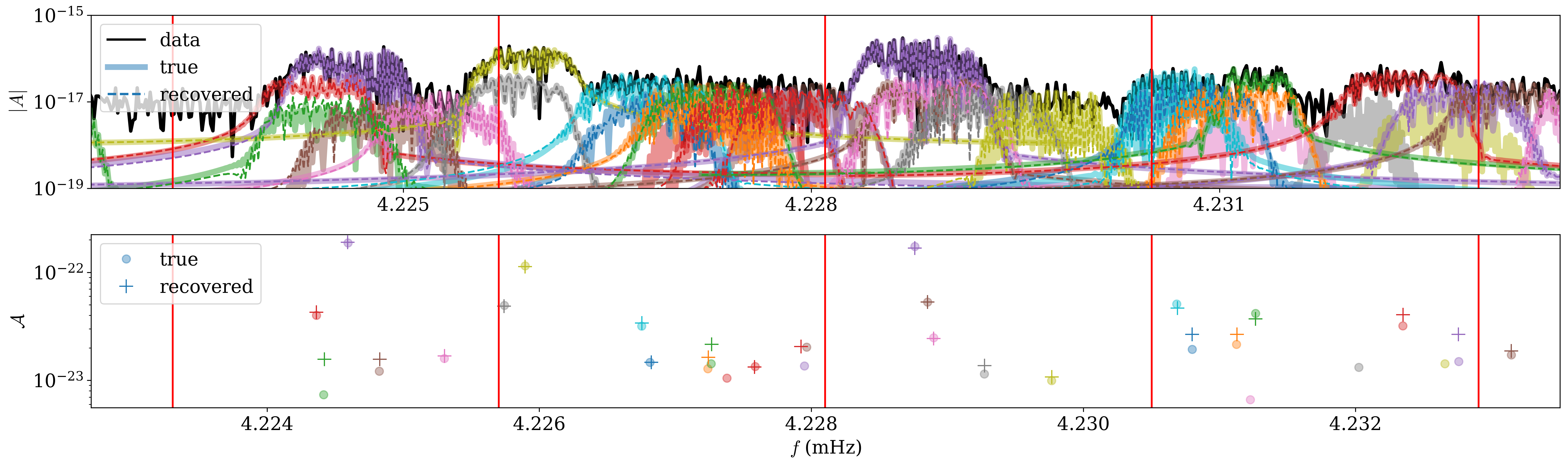}
    \caption{\textbf{Recovered GCBs by the hybrid Bayesian approach on four frequency segments.} 
    The top panel shows original data, injected signals, and recovered signals in the A TDI channel.
    The bottom panel illustrates the true and recovered values for the amplitude parameter of each source. The red vertical lines denote the boundaries of frequency segments which are divided for the convenience of extraction of GCBs considering the quasi-monochromatic feature of GCB signals.
    This figure is cited from \cite{Strub2023}.}
    \label{fig_hybrid4mHz}
\end{figure*}

The DE algorithm is a simple and efficient method for optimization problems that searches the optima by manipulating a population of potential solutions similar to PSO introduced in Section \ref{subsec_iterative}.
The DE algorithm is initialized by a population of candidate solutions drawn randomly within the search space.
Then, in following iterations, the algorithm generates new potential solutions by combining and mutating existing individuals in the population.
There are various strategies \cite{Qiang2014} for generating new solutions and updating populations, which may suit different problems and need to be determined through experimental runs. 
The populations will be continuously updated by replacing individuals with better fitness and eventually converge to the optima.
The SLSQ method is a gradient-based optimization method suits for problems with smooth and continuously differentiable likelihoods.
After the iterative subtraction search with the DE algorithm, the SLSQ method starts a global search with the initial values given by the DE algorithm, and iteratively steps towards a better solution which incorporates correlations of overlapping signals by the line search in a direction found through derivatives of the likelihood surface.
As examples, the true and recovered signals on a small frequency band are illustrated in Figure \ref{fig_hybrid4mHz}.

In the second step, The MCMC sampling algorithm is performed to obtain the posterior distribution for each identified signal.
The MCMC algorithm for individual sources is similar to methods widely used for parameter estimation in data analysis of ground-based detectors \cite{LVKCollaboration2020f,Thrane2019}, except that the data used in likelihood evaluation are the remains after the subtraction of identified signals in the first step excluding the signals to be analyzed, which can be expressed by
\begin{equation}
    d^{(i)}_{\text{posterior}} = d_{\text{original}} - \sum_{\hat{\boldsymbol{\theta}} \in \hat{\boldsymbol{\theta}}_{\text{recovered}}} h(\hat{\boldsymbol{\theta}}) + h(\hat{\boldsymbol{\theta}}_i),
\end{equation}
where $d^{(i)}_{\text{posterior}}$ is the data used in likelihood evaluation for the source $i$, and $\hat{\boldsymbol{\theta}}_i$ denotes the parameters of maximum likelihood estimation obtained in the first step for the source $i$.
One drawback is that the above MCMC sampling method for individual signals cannot account for the correlation of overlapping signals, which may lead to overoptimistic posterior distributions.
To address this, some level of residual signals are intentionally left in the data used for estimating the noise characterization. The partial residual data used to obtain the noise PSD can be expressed as
\begin{equation}
    d_{\text{partial}} = d_{\text{original}} - s_{\text{partial}}\sum_{\hat{\boldsymbol{\theta}} \in \hat{\boldsymbol{\theta}}_{\text{recovered}}} h(\hat{\boldsymbol{\theta}}),
\end{equation}
where $s_{\text{partial}}$ is a factor from the the range of $[0, 1]$ which has to be determined experientially and is set to  $s_{\text{partial}}=0.7$ in the work \cite{Strub2023}.

Various methods are used to improve the efficiency of MCMC sampling in the second step, for example, constraining the search space only within the promising region \cite{Strub2023}, or using Gaussian progress regression to model the likelihood \cite{Strub2022}.
The posterior distributions are typically concentrated within small regions of parameter space.
To relieve the computational burden, one can only explore the reduced parameter space based on the maximum likelihood estimation obtained in the first step.
The Fisher matrix is used to determine the boundaries of the reduced parameter space.
The Fisher matrix is widely used in forecast works \cite{Cutler1994,Finn1992}, which is an approximation of the Bayesian method under the assumption of high SNR and can provide estimations for parameter measurement uncertainty. 
The Fisher matrix is given by 
\begin{equation}
    F_{ij} = \left\langle \frac{\partial h(\hat{\boldsymbol{\theta}})}{\partial \theta_i},  \frac{\partial h(\hat{\boldsymbol{\theta}})}{\partial \theta_j} \right\rangle,
\end{equation}
where $\hat{\boldsymbol{\theta}}$ is the maximum likelihood estimation obtained in the first step, and the angle brackets are the inner product defined as Equation \ref{eq_inner_product}.
The inverse of the Fisher matrix presents the estimation of the covariance matrix of parameter measurement.
The boundaries of reduced parameter space for the MCMC sampling are determined by the variance of parameters estimated through the Fisher matrix.
Typically, 3-$\sigma$ regions are sufficient, while the practical settings need to be adjusted according to the features of posterior distributions in experimental runs, and the tolerance of computational burden or the desired coverage of the parameter space.

Additionally, in the procedure of MCMC sampling, the likelihood has to be evaluated a huge number of times. The computational cost for likelihood evaluation is one of the main barriers to MCMC sampling.
In the work \cite{Strub2023}, the contemporary computational hardware is used to perform the likelihood evaluation. The likelihood can be computed on GPU in massively parallel, which can significantly reduce the wall-time required for finishing the MCMC sampling \cite{Katz2022a,Katz2022b}.
On the other hand, as shown in the work \cite{Strub2022} the likelihood can be modeled by the Gaussian process regression, where the likelihood is modeled by a joint Gaussian distribution whose mean vector and covariance matrix are determined by training samples \cite{Rasmussen2005}. 
The computation of Gaussian distributions can be much faster than the likelihood defined through the inner product in Equation \ref{eq_likelihood}.
In the work \cite{Strub2022}, 1000 random samples are first drawn for training the Gaussian process regression model, and 500 samples are used for verification.
After this, The likelihood is replaced by the Gaussian process regression model in following MCMC sampling, which can reduce the computational cost of the entire sampling process.
The training and evaluating of the Gaussian process regression model are implemented through the \texttt{scikit-learn} package \cite{Pedregosa2011}.


The hybrid Bayesian approach combines the maximum likelihood estimation and the MCMC sampling.
A search of the iterative subtraction scheme is first performed to identify potential candidates which are used as the start in the subsequent global optimization in the joint parameter space for all overlapping signals.
Then the MCMC sampling algorithm is performed for each identified signal to obtain the posterior distributions.
This hybrid Bayesian approach uses the strategy of iterative subtraction to determine the source number and the point estimation for source parameters, which can avoid the computationally expensive trans-dimensional MCMC sampling.
Meanwhile, after the iterative subtraction search, the source parameters are again globally optimized around the identified candidates, which can rectify errors induced by inaccurate subtraction contamination and correlations of overlapping signals. 
However, since the MCMC sampling is performed for individual signals, only the marginalized posteriors can be obtained for each source. Incorporating additional uncertainty induced by signal overlapping requires tuning the algorithm configuration empirically.

\subsection{Utilization of machine learning techniques} \label{subsec_ml}
The development of machine learning techniques has led to revolutions in many fields including the data analysis of GWs.
{Various machine learning algorithms have been successfully used in tasks of signal identification and classification \cite{McLeod2024,Sakai2021,Chan2020}, parameter estimation \cite{Dax2021,Wildberger2023,Dax2023,Dax2021b,Green2020a,Green2020}, noise reduction \cite{Wang2023e}, waveform modeling \cite{McGinn2021,Liao2021}, etc. Comprehensive reviews can be found in \cite{Cuoco2020,Zhao2023}.}
Utilizing machine learning techniques in data analyses of space-borne detectors is also actively discussed \cite{Du2023,Zhao2024,Yun2023b,Zhang2022b,Ruan2023a}. 
A preliminary attempt of using the machine learning method to extract overlapping GCB signals is presented in the work \cite{Korsakova2024} where normalizing flows are used to build proposals for improving the convergence of MCMC sampling and offer a new method for sharing analysis results.

Normalizing flows are a class of machine learning algorithms that can model complicated distributions and are widely used in density estimation problems \cite{Papamakarios2019,Kobyzev2021,Winkler2019,Papamakarios2019a}.
Normalizing flows model a probability distribution through an invertible and differentiable transformation $f: \mathcal{X}\rightarrow \mathcal{Y}$ that can map the simple and tractable base probability distribution $P_\mathcal{X}(\boldsymbol{x})$ to the complicated target probability distribution $P_\mathcal{Y}(\boldsymbol{y})$. The target distribution can be given by
\begin{equation}
    P_\mathcal{Y}(\boldsymbol{y}) = P_\mathcal{X}\left(f^{-1}(\boldsymbol{y})\right) |\det J_{f^{-1}}(\boldsymbol{y})|
\end{equation}
where $|\det J_{f^{-1}}(\boldsymbol{y})|$ is the determinant of the Jacobian of the transformation $f^{-1}$ accounting for the change of volume elements of parameter spaces.
The transformation is parameterized by neural networks and can be composed of a sequence of transformations. Various kinds of tranformations are developed, a comprehensive review can be found in \cite{Kobyzev2021}.
In the training procedure, the random samples are drawn from the known target distribution, and the neural network is trained by mapping these samples into the simple base distribution.
In the inference procedure, the above process is performed inversely, the target distribution is obtained by transforming the random samples drawn from the base distribution through the trained neural network.
Drawing samples from the base distribution and computing the transformation can usually be much faster than computing the likelihood defined by the inner product. Once the training procedure is finished, normalizing flows can generate posterior distribution more efficiently than the MCMC sampling. 


In the work \cite{Korsakova2024}, the neural density estimation with normalizing flows is used in three different aspects.
In the first, the normalizing flows are used to build physical priors for amplitude and sky location based on population models.
The spatial distribution of GCBs mainly concentrates on the Galactic disk, rather than uniformly distributes on the whole sky.
The amplitude depends on the binary masses, distance, and orbital period which can also provided by population synthesis models \cite{Nelemans2001,Yu2010,Lamberts2019,Korol2020}.
Utilizing available information can help to constrain the parameter space and improve the efficiency of MCMC sampling.
Training with the simulated GCB catalog in the LDC2a \cite{Baghi2022}, the physical priors for amplitude and sky location can be constructed through normalizing flows.

In the second aspect, normalizing flows are used to construct proposal distributions from results given by former epoch observations.
Constructing proposal distributions from available samples to improve sampling efficiency has been proposed earlier in \cite{Falxa2023} where the density fit method of kernel density estimation (KDE) is used to build proposal distributions. However, the KDE method has drawbacks for high-dimensional problems where one needs to divide parameters into different groups and build KDE proposal distributions in low-dimensional parameter subspaces.
Normalizing flows provide an alternative way to build proposal distributions from available samples with its capability of modeling complicated distributions.
Available samples used to build proposal distributions can be obtained in two cases.
As mentioned before, the data are incrementally collected, and it is unpractical to analyze data waiting until the end of space-borne detector missions.
In reality, data analyses need to be repeatedly performed with growing data.
It is essential to fully utilize the results obtained previously when analyzing updated data.
Normalizing flows can fit the posterior distributions obtained in previous analyses, and be used to draw proposals in subsequent analyses, which is expected to have high acceptance rate and low autocorrelation of chains.
Meanwhile, for the trans-dimensional MCMC, by utilizing the capability of normalizing flows to model complicated distribution, the proposal distributions can be constructed through candidates identified by an ahead iterative subtraction procedure. 

Thirdly, the result posteriors can be published through normalizing flows.
As mentioned in Section \ref{sec_GCBs}, the number of resolvable GCBs is expected to be $\sim \mathcal{O}(10^4)$.
Moreover, due to the overlap among signals, parameters of different sources may have correlations which can not represented by marginalized posteriors for individual sources.
One may need a joint posterior for multiple sources.
Using samples to represent posteriors may have higher demands for data transfer and storage, and will be inconvenient when sharing the analysis results with the community.
Normalizing flows offer an effective way to model complicated distributions.
Therefore, the results sharing and data product publishing can be in the form of normalizing flow models trained by original posterior samples, from which users can generate samples of the identical distribution for any number needed.

In summary, normalizing flows offer a tool to fit arbitrary complicated distributions, which can be used to construct physical priors from simulated GCBs catalogs, build proposal distributions from available samples, and as a new representation of posteriors to share with the community substituting posterior samples.
Although a full-scale and end-to-end search pipeline for GCBs based on machine learning techniques is not available up to now. 
Related algorithms are demonstrated to be effective and are being incorporated into global fit pipelines as reported in the roadmap mentioned in \cite{Korsakova2024}.

\begin{table*}
    \centering
    \begin{threeparttable}
    \begin{tabular}{l|l|l|l}
    \hline\hline
    & \textbf{Solution} & \textbf{Key techniques} & \textbf{Reference} \\
    \hline
    \multirow{8}{*}{Iterative subtraction} & GBSIEVER & \makecell[l]{$\mathcal{F}$-statistic; \\ PSO.} & \cite{Zhang2021a,Zhang2022,Gao2023}  \\ \cline{2-4}
                                           & LMPSO-CV & \makecell[l]{$\mathcal{F}$-statistic; \\ PSO with special particle movement rules; \\ Tuned rules for removing false candidates.} & \cite{Gao2024} \\ \cline{2-4}
                                           & \makecell[l]{Combination of coarse template  
                                           \\ search and fine PSO search} & \makecell[l]{$\mathcal{F}$-statistic; \\ PSO with constrained search ranges; \\ Template search.} & \cite{Lu2022} \\ \hline
    \multirow{20}{*}{Global fitting}        & GBMCMC \tnote{1} & \makecell[l]{RJMCMC; \\ Parallel tempering; \\Blocked Gibbs update scheme; \\ $\mathcal{F}$-statistic proposal; \\ Multimodal proposal; \\Posterior-based proposal.} & \cite{Littenberg2020,Littenberg2023,Lackeos2023,Littenberg2024} \\ \cline{2-4}
                                           & Eryn \tnote{2} & \makecell[l]{RJMCMC; \\ Parallel tempering; \\Blocked Gibbs update scheme; \\ Ensemble sampling; \\ Delay rejection; \\Multiple try metropolis \\ Stretch-move proposal \\ Proposal built from available samples;\\ GPU acceleration.} & \cite{Karnesis2023,Katz2024} \\ \cline{2-4}
                                           & Hybrid Bayesian approach \tnote{3} & \makecell[l]{DE; \\ SLSP; \\MCMC sampling within promising regions; \\ Gaussian progress regression.} & \cite{Strub2024,Strub2022,Strub2023} \\ \cline{2-4}
                                           & Machine learning techniques & Normalizing flows & \cite{Korsakova2024} \\ \hline
    &Early efforts with MLDCs & & \cite{Arnaud2007,Babak2008b,Babak2010,Babak2008a,Vallisneri2009} \\
    \hline\hline
    \end{tabular}
    \caption{\textbf{A collection of currently available solutions for extracting GCB signals.} Relevant codes if open source are listed in the tablenote for convenience of reference. {Most equations appearing in the corresponding chapters can be found in the references listed above.}
    }
    \begin{tablenotes}
    \item[1] \url{https://github.com/tlittenberg/ldasoft}
    \item[2] \url{https://github.com/mikekatz04/Eryn}
    \item[3] \url{https://github.com/stefanstrub/LDC-GB}
    \end{tablenotes}
    \label{tab_ref_summary}
    \end{threeparttable}
\end{table*}

\section{Unresolvable sources} \label{sec_unresolvable}
As mentioned in Section \ref{sec_GCBs}, the individually resolvable GCBs are only a small fraction of the total sources. 
The remains will form a stochastic foreground and contribute to the confusion noise in the data of space-borne detectors.
On the one hand, the unresolvable GCBs play the role of noise and affect the observations of other types of sources.
On the other hand, the unresolvable sources can still provide invaluable information about the Galaxy.

\subsection{Seperation of the stochastic Galactic foreground}
The Galactic confusion noise will degrade the sensitivity of detectors and reduce the SNR of other sources \cite{Seto2004,Robson2017,Wu2023a,Liu2023}.
Additionally, due to the orbital motion of detectors, the pointing of constellations will constantly change relative to the Galactic center where GCBs are concentratedly distributed, which influences the response to the population of unresolvable GCBs and induces the modulation of the Galactic confusion noise.
Therefore the Galactic confusion noise is nonstationary, adding more challenges in data analyses of space-borne detectors \cite{Digman2022,Edwards2020}.
Furthermore, the foreground of unresolvable GCBs will blend together with the extra-galactic stochastic background of astrophysical or cosmological origin.
It is important to separate or subtract the Galactic foreground for studies of other stochastic GW background signals \cite{Flauger2021,Adams2010,Adams2014,Banagiri2021,Boileau2021,Poletti2021}.

The Galactic stochastic foreground can be separated from other components of stochastic signals or noise thanks to its distinct spectral shape and the modulation induced by the orbital motion of detectors.
For the separation, one needs to first properly model the different components.
The stochastic signals can be characterized by the cross spectrum $\langle \tilde{s}_i(f) \tilde{s}_j^*(f)\rangle$ where $i$ and $j$ denotes different TDI channels. 
Here the angle brackets denote the ensemble average.
The total stochastic signals may be contributed by three components including the confusion foreground originating from unresolvable GCBs, the stochastic background of extra-galactic origin, and the instrument noise as
\begin{equation}
    \langle \tilde{s}_i \tilde{s}_j^*\rangle= 
    \langle \tilde{h}_{i} \tilde{h}_{j}^{*}\rangle_{\text{gal}} + 
    \langle \tilde{h}_{i} \tilde{h}_{j}^{*}\rangle_{\text{extra-gal}} +
    \langle \tilde{n}_i \tilde{n}_j^*\rangle_{\text{ins}}.
\end{equation}
There are various models developed to describe these components.
For example, a typical model for extra-galactic stochastic background reads
\begin{equation}
    \langle \tilde{h}_{i} \tilde{h}_{j}^{*}\rangle_{\text{extra-gal}}= \frac{3 H_0^2 A_*}{4\pi^2 f^3} \left(\frac{f}{1 \text{mHz}}\right)^m R_{ij},
\end{equation}
where $R_{ij}$ is the response functions of detectors for different channels averaged over sky locations and polarizations, $H_0$ is the Hubble constant, $m$ is the spectral index, and $A_*$ is the amplitude at the reference frequency 1 mHz.
This power-law model is widely used in both stochastic signals of cosmological origin \cite{Bartolo2016} and astrophysical origin \cite{Abbott2019d}.
More models for extra-galactic stochastic background can be found in \cite{Flauger2021}.

Modeling instrument noise can be extremely complicated involving detailed studies about the electronic systems, optical systems, and space environments of detectors, etc.
However, in discussions of data analyses or science problems, a simplified noise model can be used, which groups noise into two components, Interferometry Metrology System (IMS) noise and acceleration noise, characterized by two quantities $S_{\text{IMS}}$ and $S_{\text{acc}}$ respectively \cite{Stas2020,Babak2021}.
Detailed derivation for noise spectral densities and correlations of different channels for different TDI generations or levels of approximation can be found in \cite{Babak2021,Flauger2021}.

For modeling the stochastic Galactic foreground, one simplified method is utilizing the analytic fitting \cite{Cornish2017a, Robson2019, Schmitz2020} which reads
\begin{equation} \label{eq_foreground_spectral}
    \langle \tilde{h}_{i} \tilde{h}_{j}^{*}\rangle_{\text{gal}} = B_* f^{-7/3} \exp\Big[-f^\alpha - \beta f \sin(\kappa f)\Big] \Big\{1 + \tanh[\gamma (f_k-f)]\Big\} R_{ij},
\end{equation}
where fitting parameter $B_*$ controls the overall amplitude, $f_k$ controls the position of the knee-like feature in the spectrum of Galactic foreground, together with $\alpha$, $\beta$, $\gamma$, and $\kappa$ describe the spectral shape of Galactic confusion foreground.
In practice, one can only vary the overall amplitude in parameter estimation while taking the values of other parameters fitted through simulations of GCB population \cite{Flauger2021}.

However, in this simplified model, the features of anisotropy and non-stationary are averaged and not incorporated. To fully exploit the distinctive features of the Galactic confusion foreground, a numerical model \cite{Adams2014} can be used.
In this numerical model, the modulated spectrum of the Galactic foreground is characterized by 17 Fourier coefficients \cite{Cornish2002,Cornish2001a,Adams2014} which are treated as free parameters varied in parameter estimation, and their prior ranges are obtained through multiple runs of GCB population simulation.
The simulations of GCB population can be based on the Galaxy model constrained by the bright and individually resolvable GCBs \cite{Adams2012}.
To properly incorporate the modulation induced by detector orbital motion, the entire data are divided into segments of week-long during which the responses of detectors to Galactic confusion foreground are considered to have no appreciable changes.



After properly modeling all components of stochastic signals, one can use Bayesian inference to obtain estimations of free parameters in models.
The likelihood is given by
\begin{equation}
    p(s|\boldsymbol{\theta}) =\prod_{n,d}\frac{1}{(2\pi)^{N/2}|C_{d,\, ij}|}\exp\Bigg(s_{d,\, i} C_{d,\, ij}^{-1} s_{d,\,j}\Bigg),
\end{equation}
which is the probability of occurrence of observed data if the stochastic signals and noise are governed by the chosen models with parameters $\boldsymbol{\theta}$.
Here, $s$ denotes the entire observed data, $\boldsymbol{\theta}$ denotes parameters required to describe models for all components of stochastic signals, $n$ labels the data samples, $d$ labels different segments used for addressing the modulation of Galactic confusion foreground, $i$ and $j$ represent the different TDI channels. $C_{d,\, ij}$ is the correlation matrix depending on chosen models for each components in stochastic signals. The product needs to run all data samples and all segments.
Posterior distributions for model parameters can be obtained through MCMC sampling, from which one can get the separation of different components of stochastic signals and noise.

Except for the stochastic background signals of extra-galactic origin, subtracting Galactic confusion foreground also benefits observations of individual sources as presented in \cite{Badger2023} where the technique referred to as dictionary learning is used to reconstruct MBHB signals with low-SNR.
The basic idea of dictionary learning \cite{Sadeghi2013} is representing a target signal $h(t)$ through a set of basis elements called dictionary $\boldsymbol{D}$ and a sparse coefficient vector $\boldsymbol{\alpha}$ as $h \sim \boldsymbol{D}\boldsymbol{\alpha}$. The dictionary can be predefined by training dataset created through signals of individual MBHBs without noise, and the coefficient vector $\boldsymbol{\alpha}$ for a given MBHB signal is searched in the presence of noise. Their combination provides a reconstruction of the signal hidden in the noisy data.
As demonstrated in the reference \cite{Badger2023}, low-SNR massive black binaries can by successfully separated in the presence of Galactic noise through this method.

\subsection{Tools for studying the Galaxy}
The old stellar population is one of the excellent tools to trace the dynamical evolution of the Galaxy \cite{Belokurov2013}.
However, these sources are usually dim and difficult to be observed in the electromagnetic band.
GCBs make up the majority of the total old stellar population and their GW signals are not affected by crowded matters in the Galaxy.
GW Observations of GCBs offer a unique tool to study the Galaxy.
Although, it has been demonstrated that the individual resolvable GCBs which have better measurement and localization can already trace the structure of the Galaxy \cite{Korol2018a,Korol2021,Wilhelm2020}.
These resolvable sources are usually more massive or nearby, which may require careful notice of potential bias \cite{Lamberts2019}.
The stochastic foreground containing the contributions of the full GCB population can also reveal various properties of the Galaxy \cite{Breivik2020a,Benacquista2006,Georgousi2022}.

In one aspect, one can extract information about the Galaxy from the spherical harmonic decomposition of the Galactic confusion foreground \cite{Breivik2020a}.
The angular power spectrum of the foreground that encodes information of the GCB spatial distribution can be obtained from the observed cross spctrum of different TDI channels by the framework presented in \cite{Taruya2005}.
It is pointed out that in \cite{Breivik2020a} the scale height of GCB distribution can be effectively constrained using the hexadecapole moment of the spherical harmonic expansion. 
Although the constraint provided by this methodology has limited sensitivity comparing to the method with resolvable GCBs, this approach may be less affected by the observation bias and provides a complementary way to measure the structure of the Galaxy.
Another approach for obtaining information about the Galaxy is through the spectral shape and amplitude of the Galactic confusion foreground \cite{Georgousi2022}.
This methodology is similar to fitting Galactic confusion foreground with the analytic model of Equation \ref{eq_foreground_spectral} mentioned in the last section.
Whereas the fitting parameters can be mapped to the physical parameters describing properties of the Galaxy. 
Thus, one can obtain information of the Galaxy like the total stellar mass as discussed in \cite{Georgousi2022} through the similar parameter estimation procedure discussed in the last section from the Galactic confusion foreground.

\section{Summary} \label{sec_summary}
Space-borne detectors will open the windows of the low GW band in the near future, which can provide new tools to explore the Universe.
In contrast to ground-based detectors whose data is noise-dominant, data from space-borne detectors will be signal-dominant where signals are more crowded and various sources are tangled together.
The abundance of sources can present plentiful invaluable information on the one hand, but also play the role of noise to disturb source detections and measurements on the other hand.
The heavily overlapping signals pose new challenges for the data analyses of space-borne detectors.

Among various source types targeted by space-borne detectors, the vast population of GCBs is likely the type having the most number to be detected.
It is expected that there are tens of millions GCBs in the mHz band, while tens of thousands massive or nearby sources among them are individually resolvable and the remains will form a stochastic foreground contributing to the confusion noise.
The GWs from GCBs are continuous signals and have the feature of quasi-monochrome. In the time domain, GCB signals with overwhelming numbers exist in the data of space-borne detectors simultaneously during the entire mission period.
In the frequency domain, although a single GCB signal is only extended within a narrow band, due to their vast number, GCB signals can still heavily overlap in frequency.
GCBs are likely the type that has the most heavy overlap and correlation.
Separation and extraction of overlapping signals focusing on GCBs may be the foundation of the ultimate data analysis pipelines for globally separating all kinds of sources.
Therefore, in the paper, we present a comprehensive review of precious efforts dedicated to the separation and extraction of GCB signals.

Current solutions for separating overlapping GCB signals can be mainly categorized into two classes, iterative subtraction and global fitting.
The strategy of iterative subtraction searches for the optimal fitting for just one signal in each iteration and the same procedure will be performed repeatedly with the remaining data after the subtraction of identified signal.
A typical iterative subtraction solution referred to as GBSIEVER employs the $\mathcal{F}$-statistic likelihood and PSO to search the optimal fitting in each step.
To address the problem of inaccurate subtraction contamination for low SNR sources, a different implementation of iterative subtraction uses a tuned particle movement rule of PSO aiming at identifying all local maxima in the likelihood surface. Then, the local maxima corresponding to false candidates induced by degeneracy of individual signals and overlapping signals are removed. Astrophysical properties of GCBs are incorporated to further filter out the low credible candidates.
Another implementation of iterative subtraction focuses on improving the efficiency in the PSO searching.
Before the fine PSO search procedure, a template search is performed to obtain the coarse estimation of existing signals.

Another strategy is global fitting which estimates the parameters of all sources together with the source number simultaneously through the full Bayesian approach.
The full Bayesian approach can provide the joint posterior distributions for multiple overlapping signals rather than the only point estimation given in the iterative subtraction scheme, which can well describe the correlation among overlapping signals and provide a straightforward way for uncertain analysis and model selection.
Posterior estimation over the parameter space with an uncertain dimension is performed through RJMCMC.
To overcome the poor convergence in the trans-dimensional MCMC sampling, various tuned proposal distributions according to the features of GCB signals and sophisticated random sampling mechanisms are adopted in different implementations.

Besides typical solutions of above two schemes, a hybrid approach that combines the point estimation of maximum likelihood and Bayesian posterior estimation is also proposed. A procedure of iterative subtraction is first performed to identify potential candidates, and a global optimization of all identified signals is performed again to relieve errors induced by correlations of overlapping signals.
Then, the MCMC sampling is performed for each found signal to obtain the marginalized posterior distributions of individual sources.
Machine learning techniques are considered as very promising tools in future GW data analyses.
Neural density estimation methods are investigated for assistance in drawing proposals and sharing the posterior results with the advantage of normalizing flows at efficiently modeling complicated distributions.

The resolvable GCBs are only a small fraction of the entire population. The remaining sources will form a stochastic foreground contributing to the confusion noise.
The Galactic confusion foreground on the one hand is a type of noise affecting the observation of other sources, on the other hand provides a powerful tool to research the structure of the Galaxy with the advantage that GWs will not be suppressed or obscured by crowded matters in the Galaxy.
Thanks to the features of the different spectral shape and the time-evolving modulation, the stochastic Galactic foreground is distinguishable with the instrument noise and the extra-galactic astrophysical or cosmological original background.
And the distinct spectral shape of Galactic foreground can also be used in the measurement of properties of the Galaxy.
The anisotropy of the Galactic foreground is associated with the stellar population distribution of the Galaxy, thus the Galactic structure can be constrained through the spherical harmonics decomposition of the Galactic foreground.

For future works, although there are already diverse end-to-end prototype pipelines for extracting overlapping GCBs, various problems still need to be addressed for actually handling real data.
For example, in current solutions the noises are usually assumed to be stationary and Gaussian, while in reality there is slow drift of instrument noise in the long period and glitches in the short period which are neither stationary nor Gaussian \cite{Spadaro2023a,Baghi2022a}. Better noise modeling may required for future works.
Besides, gaps will exist unavoidably in the data stream due to various reasons including scheduled maintenance or unplanned random events \cite{Wang2024b,Baghi2019}.
Future pipelines may need to incorporate processing of various potential defects existing in data.
Furthermore, the data from space-borne detectors blend all sources of different types, the ultimate pipelines need to be capable of addressing different kinds of overlapping signals.
New independent implementations or enchantments of current pipelines are anticipated for preparation of the launch of future space-borne GW observation missions.

\section{Acknowledgments}
The authors have no conflict of interest. W. Z. is supported by the National Key R\&D Program of China (Grant No. 2022YFC2204602 and 2021YFC2203102), Strategic Priority Research Program of the Chinese Academy of Science (Grant No. XDB0550300), the National Natural Science Foundation of China (Grant No. 12325301 and 12273035), the Science Research Grants from the China Manned Space Project (Grant No.CMS-CSST-2021-B01), the 111 Project for "Observational and Theoretical Research on Dark Matter and Dark Energy" (Grant No. B23042) and Cyrus Chun Ying Tang Foundations.
R. N. is supported in part by the National Key Research and Development Program of China Grant No.2022YFC2807303.


\bibliographystyle{elsarticle-num-names} 
\bibliography{ref}






\end{document}